\documentclass[letter,11pt]{article}
 
\usepackage{jheppub}

\usepackage[compat=1.1.0]{tikz-feynman}

\usepackage{subcaption} 
\usepackage{graphicx}
\usepackage{comment}
\usepackage{dcolumn}
\usepackage{bm}
\usepackage{float}
\usepackage{hyperref}
\usepackage{dsfont}
\usepackage{slashed}
\usepackage{color}
\usepackage{amsmath}
 \usepackage{setspace}
\usepackage[section]{placeins}
\usepackage{braket}
\usepackage{upgreek}
\usepackage[bottom]{footmisc}
\usepackage[normalem]{ulem}
\usepackage{xcolor}
\usepackage{mathrsfs}
\usepackage[leftcaption]{sidecap}
\usepackage{subcaption}

\newcommand{\nn}{\nonumber}

\def\be{\begin{eqnarray*}}
\def\ee{\end{eqnarray*}}
\def\beq{\begin{eqnarray}}
\def\eeq{\end{eqnarray}}
\def\bs{\boldsymbol} 
\def\bs{\boldsymbol}

\def\bfk{{\bs k}}

\def\bfx{{\bs x}}

\def\bfx{{\bs x}}

\newcommand{\bea}{\begin{eqnarray}}
\newcommand{\eea}{\end{eqnarray}}

\usepackage{marginnote}

\title{Dynamical Tidal response of compact stars - An EFT approach}

\author[a]{Gregory Jaurequi,}
\author[a]{Soumodeep Mitra,}
\author[a]{and Varun Vaidya}

\affiliation[a]{\footnotesize Department of Physics, University of South Dakota \\
Vermillion 57069, USA}

\emailAdd{Gregory.Jaurequi@coyotes.usd.edu}
\emailAdd{Soumodeep.Mitra@coyotes.usd.edu}
\emailAdd{Varun.Vaidya@usd.edu}

\abstract{
We apply the point particle EFT approach to a compact star to systematically compute dynamical tidal Love numbers for various non-rotating compact objects, extending the treatment of \cite{Saketh:2023bul, Saketh:2024juq}. 
We calculate the scattering amplitude in Black Hole Perturbation Theory(BHPT) for \textit{arbitrary} non-rotating compact stars using the Mano-Suzuki-Takasugi(MST) method with non zero surface reflectivity and match it with that obtained from point particle EFT order by order in the low frequency expansion. This sets up a systematic framework for extracting the static and dynamical tidal Love numbers(TLNs) to any order in the multipole expansion. In this paper, we employ the technique to compute the Next-to-Next-to Leading Order TLN upto a universal constant and its Renormalization Group equation for non-viscous Neutron stars and Neutron stars admixed with Bosonic or Fermionic dark matter. }

\begin{document}
\maketitle
\newpage

\newpage
\section{Introduction}
\label{sec:introduction}
Over the past decade the LIGO-VIRGO-KAGRA(LVK) \cite{LIGOScientific:2014pky, VIRGO:2014yos, KAGRA:2020agh, KAGRA:2021vkt,LIGOScientific:2021usb} collaboration has detected gravitational waves (GW) from various compact object mergers, like black holes (BH) and neutron stars (NS). Such detections
have provided us with a unique opportunity to test the \textit{strong-field} regime of general relativity (GR), cosmology, and high density nuclear matter etc. \cite{LIGOScientific:2016aoc, LIGOScientific:2016lio,abbott2018gw170817,Nissanke:2013fka,LIGOScientific:2017adf} in a fundamentally new way.   
With various upcoming ground (e.g. Einstein Telescope \cite{ET:2019dnz}) and space based (e.g. LISA\cite{amaro2017laser, LISA:2022kgy}) detectors our capabilities of observing such events with improved accuracy and precision will increase significantly, ushering in an era of precision gravitational wave astronomy. 
This will allow us to probe minute effects, such as deviations from the classical BH paradigm, imprints of astrophysical environments such as dark matter (DM), as well as the interior structure of compact objects.  

The orbital dynamics and gravitational waveforms of binary systems depend sensitively on the internal structure of stars which are described by an equation of state (EoS). The EoS characterizes how the density of matter that constitutes the star relates to its pressure.   
For the case of a NS, the proposed theoretical models for EoS are diverse, with large uncertainty due to a lack of knowledge of how strongly interacting matter behaves at high densities. The situation becomes more complex when one considers NS immersed in some environment, such as dark matter (DM). Such an environment is expected in extreme mass ratio inspirals (EMRIs) around galactic supermassive BHs. In such cases, accretion of DM inside a NS gives rise to a DM admixed NS \cite{Jockel:2023rrm,Cipriani:2025tga,Kain:2021hpk,Cronin:2023xzc}. This may significantly alter internal properties based on the DM mass, cross-section between DM and nuclear matter, and other factors~\cite{santos2025observational}. 

Knowledge of the EoS allows us to compute the NS tidal deformability, i.e. the tidal response of the star to an external gravitational field. This is described via a multipole expansion. The leading order term is defined as the \textit{static Tidal Love Numbers} (static TLNs), which measures the deformability of the object under time independent external tidal fields \cite{poisson2014gravity}. These have been studied extensively in the literature and have the remarkable property that they vanish exactly for BHs in GR \cite{LeTiec:2020spy,Bhatt:2023zsy,Chia:2020yla}, but are non-zero for compact objects such as NS \cite{Hinderer:2007mb}.

Static TLNs have been extensively used to probe the interior structure of NS. Greater tidal deformability results in a faster loss of orbital energy, which appears as a phase shift in the GW waveform. By determining how the NS EoS affects these TLNs, constraints on EoS have been imposed \cite{Koehn:2024gal, santos2025observational}. In recent literature, investigation into TLN and subsequent constraints on various DM models using GW observations have been extended to the arena of other compact object e.g. DM admixed NS \cite{Diedrichs:2023trk,RafieiKarkevandi:2021hcc,Giangrandi:2022wht,Mariani:2023wtv,Mukherjee:2025omu}. 
 
Most studies on TLN, especially for NS and DM admixed NS systems, (see \cite{Saes:2025jvr, Chakrabarti:2013lua,HegadeKR:2024agt} for some treatment of beyond static TLNs in NS) focus on the static regime. Although, in recent years, interest in higher order tidal responses has been growing steadily. This includes both time dependent tidal deformability given by \textit{dynamical} TLN (dTLN) and tidal dissipation numbers (TDN), corresponding to the dissipative part of the tidal response. These typically appear at higher order in the multipole expansion compared to the static TLN. This is a consequence of higher order dynamical tides increases during the late stage inspiral. In this region, with both compact objects in proximity, the time dependence of external tidal field becomes important. Here, the time scale of the internal tidal responses becomes non-negligible in comparison to orbital timescale as has been shown in \cite{Steinhoff:2016rfi,Hinderer:2016eia}. Furthermore, \cite{Pratten:2021pro} has shown that neglecting dynamical tides results in large biases in the estimation of nuclear properties of NS. On the other hand, the dissipative part, i.e. TDN (which comes at next-to leading order in the multi-pole expansion of the tidal response function for NS), is related to the viscosity of the interior material and was found to be quite small~\cite{bildsten1992tidal}. However, in recent literature, authors have shown the effect of tidal dissipation in the late-inspiral waveform to be quite important~\cite{Ripley:2023lsq,Ripley:2023qxo}.

As with NS, dynamical responses also become important for BHs and other compact objects (CO)s when modeling inspiral waveforms. While BHs in GR have vanishing static TLNs, the same is not true for dynamical TLNs (dTLN)\cite{Chakraborty:2023zed,Saketh:2023bul,Bhatt:2024yyz}. Therefore, this should be taken into account to accurately model GW waveforms. Additionally, any modifications to a standard BH, such as compact objects, BHs with environments, presence of higher dimensions, and modification of asymptotically flat GR etc. is known to give rise to non-zero TLN \cite{DeLuca:2021ite, DeLuca:2022xlz,Brito:2023pyl,Chakravarti:2018vlt,Pereniguez:2021xcj,Dey:2020pth,Cardoso:2017cfl,Nair:2024mya,Barbosa:2025uau,Franzin:2024cah}. Therefore, potential non-zero TLN observations, requiring precise GW waveform models, would signal physics beyond GR. Furthermore, non-zero BH TDNs can contribute to significant dephasing in EMRIs \cite{Datta:2024vll, Datta:2019epe}. This may be used to probe horizon-scale corrections originating from any underlying quantum gravity model \cite{Maselli:2017cmm, Nair:2022xfm}. 

Calculating the dynamical tidal response is not without challenges and subtleties. In GR, the definition of TLN suffers from ambiguity due to coordinate dependent mixing of the external tidal field and the object's response. In case of static TLNs, such issues can be avoided using analytical continuation techniques; however this method cannot be extended for dynamical TLN \cite{Chakraborty:2025wvs,Kol:2011vg,LeTiec:2020spy}. Additionally, in recent literature\cite{Katagiri:2024wbg, HegadeKR:2024agt}, authors have found an arbitrary, convention-dependent constant term entering at Next-to-Leading Order (dTLN). In practice, such terms can introduce significant biases in the estimation of NS properties. These problems can be solved by using a worldline effective field theory (wEFT) or point particle EFT approach \cite{Goldberger:2004jt, Goldberger:2006bd} (see \cite{Porto:2016pyg} for review). wEFT approach has been successfully used previously to calculate dTLN for BH in GR \cite{Ivanov:2022hlo,Saketh:2023bul, Wang:2026qst} and higher -curvature theories, as well as for static TLn for NS\cite{Saketh:2024juq}.

In this approach, one approximates the CO as a point particle at leading order, while the finite size effects are encoded via higher dimensional operators. Here, the Wilsonian coefficients represent the tidal response of the object. These coefficients can then be obtained by matching some suitable observable calculated from both EFT and full theory GR.
In \cite{Saketh:2023bul}, the authors have obtained the dynamical tidal response of a rotating BH to all orders in spin and to Next-to-Leading order (NLO) in  the small frequency approximation. In \cite{Saketh:2024juq}, the authors first used this technique to obtain leading order TLN and TDN for NS.  In this paper, we extend previous computations on static and dTLN for a class of COs. We first improve the result in \cite{Saketh:2024juq} to NNLO, computing the static and lowest  dynamical TLN for a NS with no viscosity (hence zero dissipation). We then expand the regime of the previous work to include DM admixed NS systems. Once again, we calculate up to NNLO and comment on the effect of various DM properties and explore the Renormalization Group evolution for these dTLNs. Recently, following our work, various other groups also attempted to calculate dTLN of pure NS system, using an EFT approach and investigated other aspects, such as relations with Quasi-normal modes \cite{Saketh:2026trm} and detection prospects \cite{Apostolidis:2026qsg}. 

This paper is structured as follows: In Section~\ref{EFT} we review the wEFT method, define the TLNs and TDNs, and compute the graviton-CO scattering amplitude used as the observable for matching. In Section~\ref{sec:BHPT} we calculate the same observable in full GR using MST formalism for a CO with non-zero reflectivity. We then solve the interior solution of the compact object for both a NS and NS+DM system to derive the surface reflectivity of the star in terms of its interior properties in Section~\ref{sec:Interior soln}. We then use these results to match the scattering amplitude to the EFT result in order to derive the tidal responses of various compact objects in Section~\ref{sec:Matching}. We present our conclusions in Section~\ref{Conc}.

Throughout the paper we will use $c=1$ units and the mostly plus metric signature $(-,+,+,+)$ . Greek indices denote components $0,...,3 $ and Latin indices will denote spacelike components. 


\section{Point Particle EFT}
\label{EFT}
Let us first briefly describe the technique of obtaining the TLN with an EFT approach. In point particle or wEFT formalism, the CO, when probed at a length scale $r$ much larger than its size $R$, can be effectively treated as a point particle at leading order in the ratio $r/R$. The finite size effects, carrying information about the interior structure, can be incorporated via higher order terms via a multipole expansion. Static and dynamical TLN and TDN in this formalism can be defined as the Wilsonian coefficients of these higher order operators in a gauge invariant manner. However, to extract the values of the coefficients, one needs to match the EFT with the full UV-complete theory, namely black hole perturbation theory (BHPT). This matching can then be accomplished by comparing gauge-invariant observables calculated from EFT and BHPT, such as the scattering amplitude of a long wavelength GW impinging on the CO. To do this matching explicitly, we first calculate the scattering amplitude in both wEFT formalism and BHPT, then match them order by order in the expansion parameter $\epsilon = 2 GM \omega$. Here, $\omega$ is the frequency of the impinging wave and $M$ the mass of the CO. For the case of NS or DM admixed NS systems, the scattering amplitude obtained from BHPT will contain information about the interior structure of the CO in terms of a surface reflectivity term, which we will explicitly obtain by solving Einstein's field equations in the interior of the CO. 

\subsection{The EFT action}

We start with a brief review of the point particle EFT formalism for compact objects \cite{Goldberger:2004jt,Goldberger:2006bd}. At leading order, the dynamics is described by a massive point particle action coupled to gravity 
\bea
S_{p.p} = -M\int d\tau = -M \int d\lambda \sqrt{-\frac{dx^{\mu}}{d\lambda}\frac{dx^{\nu}}{d\lambda}g_{\mu\nu}},
\eea
where  $\tau$ is the proper time and $\lambda$ can be any monotonically increasing affine parameter. The leading finite size effects are introduced through higher dimensional operators that satisfy symmetry constraints, i.e. diffeomorphism and reparameterization invariance. In a vacuum background, the Ricci tensor is zero, hence the leading non-trivial object is the Weyl tensor, the traceless part of the Riemann tensor. For a particle without spin degrees of freedom, the only non-trivial tensor is the 4-velocity $u^{\mu} = dx^{\mu}/d\tau$. Hence the lowest order action beyond the point particle that couples the particle to gravity can be written as 
\bea
S= S_{p.p}+ \int d\tau\Big[ c_E E^{\mu \nu}E_{\mu \nu}+c_BB^{\mu \nu}B_{\mu \nu}\Big]
\eea
where 
\bea
E^{\mu \nu} =C^{\mu \rho \nu \sigma}u_{\rho}u_{\sigma},  \ \ B^{\mu \nu} =(1/2)\epsilon_{\gamma(\mu}^{\alpha \beta}C_{\alpha \beta|\nu)\delta}u^{\gamma} u^{\delta}
\eea
where $C_{\mu\nu\rho\sigma}$ is the Weyl tensor. Note that the $E \cdot B$ type term is forbidden, since with a non-spinning point particle in GR, parity is a good symmetry. 
For the case of a static particle with $u^{\mu} =(1,0,0,0)$, given that the tidal fields $E^{\mu \nu}$, $B^{\mu \nu}$ are orthogonal to the 4-velocity, only their projection orthogonal to the direction of 4-velocity are relevant. Since $[E^2] =[B^2] = 4$, these operators are suppressed by $1/r^4$. Hence to define a dimensionless Wilson coefficient, we can factor out terms which scale as the size of a black hole with mass M, namely $r_H \sim GM$
\bea
S= S_{p.p}- M(GM)^4\int d\tau \Big[\Lambda_E E^{ij}E_{ij}+\Lambda_BB^{ij}B_{ij}\Big]
\eea
We can extend this action to account for higher order effects in two ways. First, taking higher order spatial derivatives\footnote{The projection along the `spatial' direction is defined as $P^{\mu}_{\ \nu}= \delta^{\mu}_{ \ \nu}+u^{\mu}u_{\nu}$} gives access to higher multiple moments and hence higher static TLNs. Secondly, higher order proper time derivatives will give \textit{dynamical} TLNs. The action can then be written as 
\bea
S= S_{p.p}-M(GM)^4 \sum_{n=0}^{\infty}\sum_{l=2}^{\infty}\int d\tau \Bigg[(-1)^n(GM)^n\Lambda_{\omega^n}^{E,l}\frac{d^n\nabla^{i_3}..\nabla^{i_l}E^{ij}}{d\tau^n}\nabla_{i_3}..\nabla_{i_l}E_{ij} + E \leftrightarrow B\Bigg]
\eea
 Note that this series includes operators that are odd under time reversal, signifying an irreversibility of dynamics, and hence encoding absorption/dissipation. Higher order operators with proper time derivatives are suppressed by the parameter $GM\omega$ which, along with $R/r$, forms another expansion parameter of our EFT. This indicates that in the EFT we do not excite any internal degrees of freedom of the star. 
 \subsection{Observable for matching to microscopic theory}
 The Wilson coefficients $\Lambda^{E/B,l}_{\omega^n}$ may be obtained either by comparing the predictions of this EFT for a specific observable with data, or by an order by order expansion in $R/r, GM\omega$ of an explicit calculation if the microscopic theory is known. In this paper, we will do the latter. The observable chosen is the amplitude for scattering a low frequency ($GM\omega \ll 1$) graviton off our heavy static compact object of mass M
 \bea
&& _{\text{out}}\langle \bfk',h'|\bfk,h\rangle_{\text{in}} \equiv \langle \bfk',h'|S|\bfk, h\rangle = \langle \bfk',h'|\mathbf{1}+iT|\bfk, h\rangle \nn\\
 &=& 2\omega(2\pi)^3 \delta^3(\bfk -\bfk')\delta_{hh'}+ 2\pi\delta(\omega-\omega')\times i\mathcal{M}(\omega, \bfk\rightarrow \bfk', h \rightarrow h') 
 \eea
 where $\bfk, \bfk'$ are the incoming and outgoing momenta of a graviton with helicity $h$ and $h^{'}$ respectively. 
The scattering amplitude is proportional to $\delta(\omega-\omega')$ which conserves the energy of the graviton. This follows if we assume the 4-velocity $u^{\mu}$ to be independent of $\tau$, which effectively is taking the limit $M \rightarrow \infty$. In principle, the 4-velocity fluctuates, and recoil effects are needed to accurately compute the scattering amplitude. However, if we are only interested in the tidal Love numbers and dissipation numbers, then we may ignore recoil in both the full theory and the effective theory, i.e. effectively take $M \rightarrow \infty$( while keeping $GM\omega \ll 1$) before matching. This should suppress all recoil effects which will appear as powers of 1/M.
 For the purpose of matching, we need the scattering amplitude in the spherical basis, which can be computed as 
  \bea
 && \mathcal{A} (\omega ,l,m, h \rightarrow \omega ,l',m',h') =\nn\\
 &&\int \frac{d^3\bfk_1 d^3\bfk_2}{(2\pi)^6 \times 4|\bfk_1||\bfk_2|} \sum_{h_1,h_2} \langle \omega, l', m', h'| \bfk_2,h_2\rangle \mathcal{M}(\omega, \bfk_1 \rightarrow \bfk_2,  h_1 \rightarrow h_2)\langle \bfk_1, h_1 |\omega, l, m, h \rangle
 \eea
 The overlap of the plane wave with the spherical basis is given as 
 \bea
 \langle \omega, l, m, h|\bfk, h\rangle = (2\pi)^2 \sqrt{\frac{2l+1}{2\pi \omega}} \delta(\omega - |\bfk|)\delta_{hh'} D^l_{mh}(\phi, \theta)
 \eea
 with 
 \bea
   D^l_{mh}(\phi, \theta) = (-1)^m \sqrt{\frac{4\pi}{2l+1}} \ \ _hY_{l,-m}(\phi, \theta)
 \eea
where $_hY_{l,-m}(\phi, \theta)$ are the spin weighted spherical harmonics. We see that the amplitude $\mathcal{A}$ is dimensionless and can be related to the absorption probability and elastic scattering phase shift in the usual manner 
 \bea
 i\mathcal{A}= 1- \eta_{lm} e^{2i \delta_{lm}}
 \label{eq:Amp}
\eea
 where $1-\eta_{lm}^2$ is the absorption probability.

\subsection{The graviton scattering amplitude}
Using our action, we can now compute the scattering amplitude $i\mathcal{M}$ order by order in perturbation theory. We only consider the tidal operators, while ignoring the contribution from the point particle EFT. This amounts to ignoring the graviton scattering from the long range gravitational potential of the point object. This follows from the near-far zone factorization technique introduced in \cite{Ivanov:2022qqt}, and in subsequent sections we will explain this technique and its shortcomings. In the transverse traceless or radiation gauge, the tidal field at leading order in the metric perturbation $h_{\mu \nu}$ is simply $E_{ij}(x^{\mu}(\tau)) = -1/2\ddot{h}_{ij}(x^{\mu}(\tau))$. For a static object, using $\langle 0|h_{ij}(\bfx)|\bfk,h\rangle = 1/M_{\text{pl}}\exp{\left(-ik \cdot x\right)}\epsilon_{ij}^h(\bfk)$, we obtain 
\bea 
 i\mathcal{M}^{(0)}(\bfk_{\text{in}} \rightarrow \bfk_{\text{out}}, h\rightarrow h') = -i\frac{\omega^4}{4M_{\text{pl}}^2}\Lambda^E \epsilon_h^{ij}(\bfk_{\text{in}}) \epsilon^{*ij}_{h'}(\bfk_{\text{out}})M(GM)^4 + \text{magnetic}
 \eea

 which then allows us to write the amplitude in the spherical basis as
\bea 
 i\mathcal{A}^{(0)}(\omega, l=2,m,h \rightarrow l=2, m, h)= -i\frac{\omega^5}{40 M^2_{\text{pl}}\pi}M(GM)^4\Lambda_{\omega^0}^E +\text{magnetic}
\eea
where $M_{pl}=\sqrt{\frac{\hbar }{32\pi G}}$.
Confining ourselves to $l=2$, and going to higher order in $\omega$, we gain access to the NLO tidal responses,  
\bea 
 i\mathcal{A}^{(1)}(\omega, l=2,m,h \rightarrow l=2, m, h)= \frac{\omega^6}{40 M^2_{\text{pl}}\pi}M(GM)^5(H_{\omega^1}^E)+\text{magnetic}
\eea\
where $H_{\omega^1}^E \equiv \Lambda_{\omega^1}^E$ is the leading tidal dissipation number. 
In order to go up to NNLO, we also need the next order tidal correction which is simply 
\bea 
 \label{secondOrderEFT}
 i\mathcal{A}^{(2)}(\omega, l=2,m,h \rightarrow l=2, m, h)= -i\frac{\omega^7}{40 M^2_{\text{pl}}\pi}M(GM)^6(\Lambda_{\omega^2}^E)+\text{magnetic}
\eea
Together, the scattering amplitude with tidal operators is given by the following diagram 
\begin{equation}
\begin{aligned}
       & \quad     \vcenter{\hbox{\begin{tikzpicture}[scale=0.7]
        \begin{feynman}
            \vertex (i) at (0,0);
            \vertex (e) at (0,3);
            \node[circle, draw=green, fill = green, scale=0.5] (w1) at (0, 1.0);
            \node[circle, draw=green, fill = green, scale=0.5] (w3) at (0, 2.0);
            \vertex(f1) at (2.0,2.8);
            \vertex (fs) at (2.0,0.2);
            \vertex (f3) at (1.0, 0.70);

            \diagram*{
                (i) -- [double, double, thick] (w1),
                (w1) -- [dashed, green, ultra thick,edge label = $\Lambda^{E/B}$] (w3),
                (w3) -- [double, double, thick] (e),
                (f1) -- [boson, blue, ultra thick] (w3),
                (fs) -- [boson, blue, ultra thick] (w1),
            };
        \end{feynman}
    \end{tikzpicture}}}
\end{aligned}
\end{equation}
where $\Lambda^{E/B}$ is a short hand for tidal operators up to NNLO order. The \textcolor{blue}{blue} wavy lines represents incoming and outgoing graviton, with the solid double lines represent the CO worldline. Writing the amplitude explicitly:
\begin{equation}
\label{response}
    i\mathcal{A}_{\rm{0PM}} = -i \frac{\omega^5}{40 \pi M^2_{\text{pl}}} M (GM)^4 \mathcal{F}^E(\omega) + \text{magnetic}
\end{equation}
Where we have defined the response function $\mathcal{F}^E(\omega) = \Lambda_{\omega^0}^E + i (GM\omega) H_{\omega^1}^E+ (GM\omega)^2 \Lambda_{\omega^2}^E $ and $H^E_{\omega^1} \equiv \Lambda_{\omega^1}^E$.

\subsection{Tail effect and RG running}
The results of previous section are not complete however, as we have neglected tail effects arising from long range nature of GR in the higher order scattering amplitude. These are essentially loop corrections to the tree level diagrams considered so far. We can incorporate these pieces as Post-Minkowskian corrections to the diagrams shown above. Below we briefly describe the procedure and direct readers to \cite{Ivanov:2022hlo, Goldberger:2009qd,Porto:2012as} for further reading. 

\begin{figure}[b]
    \centering
    \includegraphics[width=0.1\linewidth]{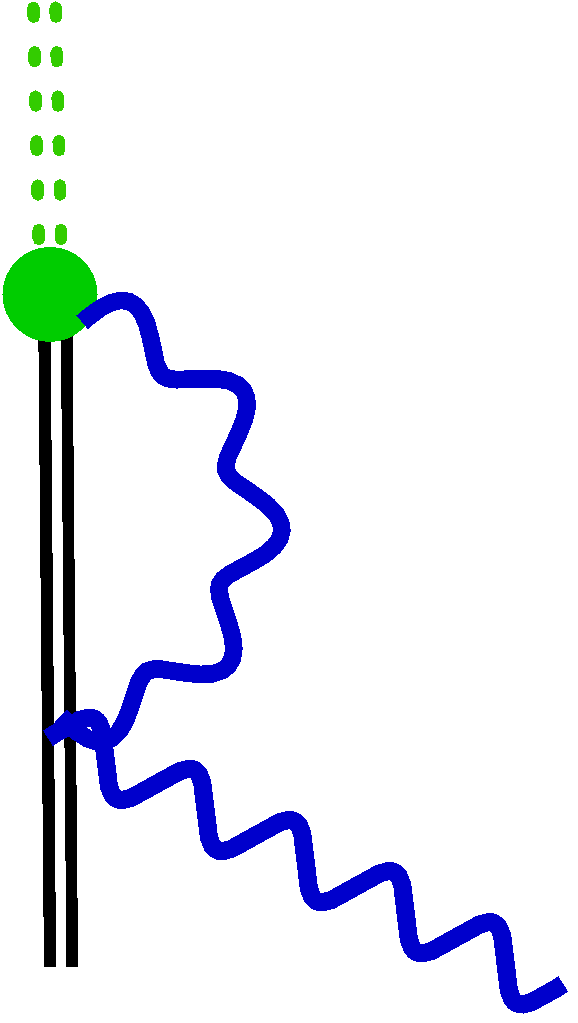}
    \caption{Seagull diagram with a possible contribution at 1PM.}
    \label{fig:seagull}
\end{figure}

We first consider the 1-PM correction manifesting as a 1-loop correction to the scattering amplitude and given by the diagrams,
\begin{equation}
\begin{aligned}
       & \quad     \vcenter{\hbox{\begin{tikzpicture}[scale=0.7]
        \begin{feynman}
            \vertex (i) at (0,0);
            \vertex (e) at (0,3);
            \node[circle, draw=green, fill = green, scale=0.5] (w1) at (0, 1.0);
            \node[circle, draw=green, fill = green, scale=0.5] (w3) at (0, 2.0);
            \vertex(f1) at (2.0,2.8);
            \vertex (fs) at (2.0,0.2); 
            \vertex[label= left:$M$] (w2) at (0,0.25);
            \vertex (f3) at (1.0, 0.70);

            \diagram*{
                (i) -- [double, double, thick] (w1),
                (w1) -- [dashed, green, ultra thick,edge label = $\lambda^{E/B}$] (w3),
                (w3) -- [double, double, thick] (e),
                (w2) -- [boson, blue, ultra thick] (f3),
                (f1) -- [boson, blue, ultra thick] (w3),
                (fs) -- [boson, blue, ultra thick] (w1),
            };
        \end{feynman}
    \end{tikzpicture}}}
    + 
                \vcenter{\hbox{\begin{tikzpicture}[scale=0.7]
        \begin{feynman}
            \vertex (i) at (0,0);
            \vertex (e) at (0,3);
            \node[circle, draw=green, fill = green, scale=0.5] (w1) at (0, 1.0);
            \node[circle, draw=green, fill = green, scale=0.5] (w3) at (0, 2.0);
            \vertex(f1) at (2.0,2.8);
            \vertex (fs) at (2.0,0.2); 
            \vertex[label= left:$M$] (w2) at (0,2.75);
            \vertex (f3) at (1.0, 2.4);

            \diagram*{
                (i) -- [double, double, thick] (w1),
                (w1) -- [dashed, green, ultra thick,edge label = $\lambda^{E/B}$] (w3),
                (w3) -- [double, double, thick] (e),
                (f1) -- [boson, blue, ultra thick] (w3),
                (fs) -- [boson, blue, ultra thick] (w1),
                (w2) -- [boson, blue, ultra thick] (f3)
            };
        \end{feynman}
    \end{tikzpicture}}} \\
    & = \vcenter{\hbox{\begin{tikzpicture}[scale=0.7]
        \begin{feynman}
            \vertex (i) at (0,0);
            \vertex (e) at (0,3);
            \node[circle, draw=green, fill = green, scale=0.5] (w1) at (0, 1.0);
            \node[circle, draw=green, fill = green, scale=0.5] (w2) at (0, 2.0);
            \vertex (f1) at (1.5,2.8) {};
            \vertex (fs) at (1.5,0.2) {};

            \diagram*{
                (i) -- [double, double, thick] (w1),
                (w1) -- [dashed, green, ultra thick,edge label = $\lambda^{E/B}$] (w2),
                (w2) -- [double, double, thick] (e),
                (f1) -- [boson, blue, ultra thick] (w2),
                (fs) -- [boson, blue, ultra thick] (w1)
            };
        \end{feynman}
    \end{tikzpicture}}} {\times 2 (16 \pi G M \omega^2) \int \frac{d^{d-1} \boldsymbol{q}}{(2\pi)^{d-1}} \frac{1}{\boldsymbol{q}^2} \frac{1}{(\omega )^2 - (\boldsymbol{q} + \boldsymbol{k})^2} }
    \end{aligned}
\end{equation}
where $\lambda^E/B$ is a short hand for tidal operators up to NNLO order. 
The integration can be explicitly solved, giving the 1PM correction to scattering amplitude, as 
\begin{equation}
\label{A_1PM_full}
    i \mathcal{A}_{1\rm{PM}} =  i \mathcal{A}_{0\rm{PM}} \times \left[ 2 G M \omega~ \rm{sign}(\omega) \pi + i (2 GM \omega) \left(\frac{2}{(d-4)_{\rm{IR}}} + \gamma_E -\frac{11}{6}+log\frac{\omega^2}{\pi \mu^2} \right) \right] .
\end{equation} 
We work in the MS bar scheme and see explicitly that the imaginary part has a logarithmic IR divergence regulated by the scale $\mu$. However, as shown in the references above, this imaginary part of the diagram contributes to a universal phase which never contributes to any observable computed in the EFT, and therefore can be dropped out. On the other hand, we see, due to the real part, the lower order terms in the scattering amplitude will start to contribute at one order higher (in $\omega$) and therefore we can replace the response function in \eqref{response} with 1-PM corrected response function as,  
\begin{equation}
    \mathcal{F}_{1\rm{PM}}^E(\omega) = \mathcal{F}^E(\omega) \times (1 + \textcolor{orange} {2GM\omega \pi})
\end{equation} 
We will use \textcolor{orange}{\textit{orange}} color terms to identify all the 1-PM tail corrections in the EFT and BHPT calculation. We also have another possible correction at this order from the two graviton vertex in the point particle action Fig.\ref{fig:seagull}.
To compute this diagram we need the 2 graviton point particle vertex, this gives the vertex
   \begin{eqnarray}
       V_{pp}^{\mu \nu, \alpha \beta}= 8\pi i GM u^{\mu}u^{\nu}u^{\alpha}u^{\beta}
\end{eqnarray}
and the graviton propagator in the harmonic or de Donder gauge
\begin{eqnarray}
    D_{\mu \nu, \alpha\beta} = \frac{i P_{\mu \nu, \alpha\beta}}{q^2+i\epsilon}
\end{eqnarray}
\begin{eqnarray}
    P_{\rho \sigma,\alpha \beta}= \frac{1}{2}\left(\eta_{\rho \alpha} \eta_{\sigma \beta}+ \eta_{\rho \beta}\eta_{\sigma \alpha}- \frac{2}{D-2}\eta_{\rho \sigma}\eta_{\alpha \beta}\right)
\end{eqnarray}
The contraction of these two vertices gives  
\begin{eqnarray}
    V_{pp}^{\mu\nu, \alpha \beta}D_{\alpha \beta, \rho \sigma} = 4\pi i GM u^{\mu}u^{\nu}\left( u_{\rho}u_{\sigma}-\frac{1}{D-2}\eta_{\rho \sigma}\right)
\end{eqnarray}

However, we immediately note that given $u^{\mu} =(1,0,0,1)$ the diagram in Fig.~\ref{fig:seagull} is proportional $\epsilon^{00}(\bfk)$ which is zero for physical graviton helicity states. Therefore, this contribution vanishes and we can infer that all diagrams with an insertion of the two graviton point particle vertex connected to an external graviton will vanish as well. 

Since we wish to compute up to $\Lambda_{\omega^2}$, we also need to consider 2-PM corrections. At this order, the corrections are given by the diagrams, 
\begin{equation}
\begin{aligned}
    & \quad
    2 \times \left [ \vcenter{\hbox{\begin{tikzpicture}[scale=0.8]
        \begin{feynman}
            \vertex (i) at (0,0);
            \vertex (e) at (0,2.0);
            \node[circle, draw=green, fill = green, scale=0.5, label=left:$\lambda^{E/B}$] (w1) at (0, 1.0);
            \vertex (fs) at (1.5,0.2) {};
            \vertex[label= left:$M$] (w2) at (0,0.5);
            \vertex (f1) at (0.75, 0.8) {};
             \vertex[label= left:$M$] (w3) at (0,0.0);

            \diagram*{
                (i) -- [double, double, thick] (w1),
                (w1) -- [dashed, double, double,green, thick] (e),
                (fs) -- [boson,blue, ultra thick] (w1),
                (w2) -- [boson, blue, ultra thick] (f1),
                (w3) -- [boson, blue, ultra thick] (f1),
            };
        \end{feynman}
    \end{tikzpicture}}}
    + \qquad
    \vcenter{\hbox{\begin{tikzpicture}[scale=0.8]
        \begin{feynman}
            \vertex (i) at (0,0);
            \vertex (e) at (0,2.0);
            \node[circle, draw=green, fill = green, scale=0.5, label=left:$\lambda^{E/B}$] (w1) at (0, 1.0);
            \vertex (fs) at (1.5,0.2) {};
            \vertex[label= left:$M$] (w2) at (0,0.5);
            \vertex (f1) at (0.75, 0.8) {};
             \vertex[label= left:$M$] (w3) at (0,0.0);
             \vertex (ib) at (0.4,0.45);

            \diagram*{
                (i) -- [double, double, thick] (w1),
                (w1) -- [dashed, double, double, green, thick] (e),
                (fs) -- [boson,blue, ultra thick] (w1),
                (w2) -- [boson, blue, ultra thick] (ib),
                (w3) -- [boson, blue, ultra thick] (ib),
                (ib) -- [boson, blue, ultra thick] (f1),
                };
        \end{feynman}
    \end{tikzpicture}}}
    + \qquad
    \vcenter{\hbox{\begin{tikzpicture}[scale=0.8]
        \begin{feynman}
            \vertex (i) at (0,0);
            \vertex (e) at (0,2.0);
            \node[circle, draw=green, fill = green, scale=0.5, label=left:$\lambda^{E/B}$] (w1) at (0, 1.0);
            \vertex (fs) at (1.7,0.25) {};
             \vertex (tfs) at (1.2,0.6);
            \vertex[label= left:$M$] (w2) at (0,0.5);
            \vertex (f1) at (0.75, 0.8);
             \vertex[label= left:$M$] (w3) at (0,0.0);

            \diagram*{
                (i) -- [double, double, thick] (w1),
                (w1) -- [dashed, double, double,green, thick] (e),
                (tfs) -- [boson,blue, ultra thick] (w1),
                (w2) -- [boson, blue, ultra thick] (f1),
                (w3) -- [boson, blue, ultra thick] (tfs),
                (tfs) -- [boson,blue, ultra thick] (fs),
                };
        \end{feynman}
    \end{tikzpicture}}} \right]
    + \qquad
    \vcenter{\hbox{\begin{tikzpicture}[scale=0.7]
        \begin{feynman}
            \vertex (i) at (0,0);
            \vertex (e) at (0,3);
            \node[circle, draw=green, fill = green, scale=0.5] (w1) at (0, 1.0);
            \node[circle, draw=green, fill = green, scale=0.5] (w3) at (0, 2.0);
            \vertex(f1) at (2.0,2.8);
            \vertex (fs) at (2.0,0.2); 
            \vertex[label= left:$M$] (w2) at (0,0.25);
            \vertex (f3) at (1.0, 0.70);
            \vertex[label= left:$M$] (w4) at (0,2.75);
            \vertex (f4) at (1.0, 2.4);

            \diagram*{
                (i) -- [double, double, thick] (w1),
                (w1) -- [dashed, green, ultra thick,edge label = $\lambda^{E/B}$] (w3),
                (w3) -- [double, double, thick] (e),
                (w2) -- [boson, blue, ultra thick] (f3),
                (f1) -- [boson, blue, ultra thick] (w3),
                (fs) -- [boson, blue, ultra thick] (w1),
                (w4) -- [boson, blue, ultra thick] (f4)
            };
        \end{feynman}
    \end{tikzpicture}}}
\end{aligned}
\end{equation}

\begin{figure}
    \centering
    \includegraphics[width=0.4\linewidth]{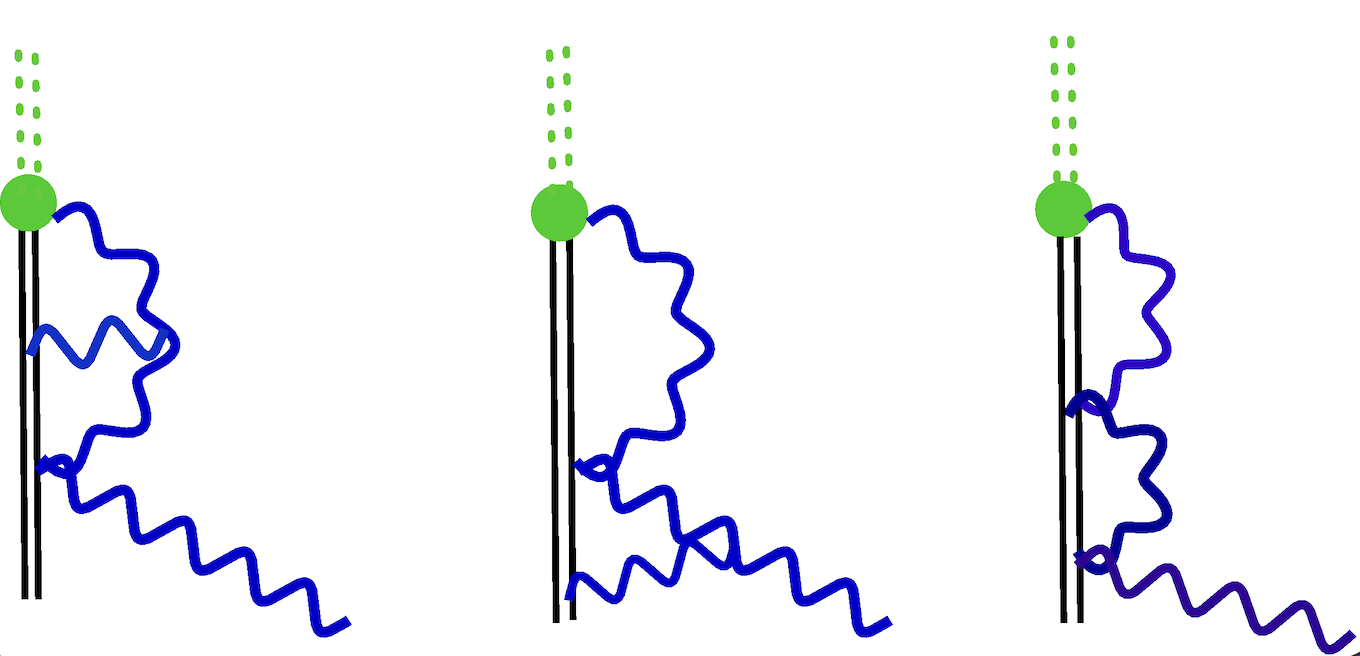}
    \caption{Additional diagrams at 2 PM involving the point particle two graviton vertex. However, all of these contributions vanish.}
    \label{fig:2loop}
\end{figure}

These corrections include IR divergences as well as UV divergences that need to be renormalized. Let us consider the first three diagrams. Using the detailed calculation in \cite{Goldberger:2009qd}, 
\begin{align}
\label{A_2PM_a-c}
    i \mathcal{A}^{(a+b+c)}_{\rm{2PM}} = i \mathcal{A_{\rm{0PM}}} \times& (G M | \omega |)^2 \times \left(-\frac{\omega^2 }{\pi \mu^2} e^{\gamma_E}\right)^{(d-4) }\times  \\ \nn  
    &\left( - \frac{2}{(d-4)^2_{\rm{IR}}} + \frac{11}{3}\frac{1}{(d-4)_{\rm{IR}}} - \frac{107}{105} \frac{1}{(d-4)_{\rm{UV}}} - \frac{7 \pi^2}{12} - \frac{1777}{14700}   \right)  
\end{align}
On the other hand, the last diagram can be written in terms of the 1PM loop corrections as, 
\begin{equation}
\begin{aligned}
    \vcenter{\hbox{\begin{tikzpicture}[scale=0.7]
        \begin{feynman}
            \vertex (i) at (0,0);
            \vertex (e) at (0,3);
            \node[circle, draw=green, fill = green, scale=0.5] (w1) at (0, 1.0);
            \node[circle, draw=green, fill = green, scale=0.5] (w3) at (0, 2.0);
            \vertex(f1) at (2.0,2.8);
            \vertex (fs) at (2.0,0.2); 
            \vertex[label= left:$M$] (w2) at (0,0.25);
            \vertex (f3) at (1.0, 0.70);
            \vertex[label= left:$M$] (w4) at (0,2.75);
            \vertex (f4) at (1.0, 2.4);

            \diagram*{
                (i) -- [double, double, thick] (w1),
                (w1) -- [dashed, green, ultra thick,edge label = $\lambda^{E/B}$] (w3),
                (w3) -- [double, double, thick] (e),
                (w2) -- [boson, blue, ultra thick] (f3),
                (f1) -- [boson, blue, ultra thick] (w3),
                (fs) -- [boson, blue, ultra thick] (w1),
                (w4) -- [boson, blue, ultra thick] (f4)
            };
        \end{feynman}
    \end{tikzpicture}}}
     = \qquad
     \vcenter{\hbox{\begin{tikzpicture}[scale=0.7]
        \begin{feynman}
            \vertex (i) at (0,0);
            \vertex (e) at (0,3);
            \node[circle, draw=green, fill = green, scale=0.5] (w1) at (0, 1.0);
            \node[circle, draw=green, fill = green, scale=0.5] (w3) at (0, 2.0);
            \vertex(f1) at (2.0,2.8);
            \vertex (fs) at (2.0,0.2); 
            \vertex[label= left:$M$] (w2) at (0,0.25);
            \vertex (f3) at (1.0, 0.70);

            \diagram*{
                (i) -- [double, double, thick] (w1),
                (w1) -- [dashed, green, ultra thick,edge label = $\lambda^{E/B}$] (w3),
                (w3) -- [double, double, thick] (e),
                (w2) -- [boson, blue, ultra thick] (f3),
                (f1) -- [boson, blue, ultra thick] (w3),
                (fs) -- [boson, blue, ultra thick] (w1),
            };
        \end{feynman}
    \end{tikzpicture}}}
    \times 
                \vcenter{\hbox{\begin{tikzpicture}[scale=0.7]
        \begin{feynman}
            \vertex (i) at (0,0);
            \vertex (e) at (0,3);
            \node[circle, draw=green, fill = green, scale=0.5] (w1) at (0, 1.0);
            \node[circle, draw=green, fill = green, scale=0.5] (w3) at (0, 2.0);
            \vertex(f1) at (2.0,2.8);
            \vertex (fs) at (2.0,0.2); 
            \vertex[label= left:$M$] (w2) at (0,2.75);
            \vertex (f3) at (1.0, 2.4);

            \diagram*{
                (i) -- [double, double, thick] (w1),
                (w1) -- [dashed, green, ultra thick,edge label = $\lambda^{E/B}$] (w3),
                (w3) -- [double, double, thick] (e),
                (f1) -- [boson, blue, ultra thick] (w3),
                (fs) -- [boson, blue, ultra thick] (w1),
                (w2) -- [boson, blue, ultra thick] (f3)
            };
        \end{feynman}
    \end{tikzpicture}}} 
\end{aligned}
\end{equation}
which then gives, 
\begin{align}
\label{A_2PM_d}
    i \mathcal{A}^{(d)}_{\rm{2PM}} = i \mathcal{A_{\rm{0PM}}}  & \times (-G M | \omega |)^2 \times \left(-\frac{\omega^2+i \epsilon }{\pi \mu^2} e^{\gamma_E}\right)^{(d-4)} \\ \nn & \left[ \frac{4}{(d-4)^2_{\rm{IR}}} - \frac{22}{3}\frac{1}{(d-4)_{\rm{IR}}} + \left(\frac{11}{6}\right)^2 + 4 \left(\frac{\pi^2}{8} + \frac{203}{72}\right) - \frac{11}{3}\left(\frac{\pi^2}{8}+\frac{203}{72}\right) (d-4)_{\rm{IR}}  \right]
\end{align}
We also have corrections at this order from the 2 graviton point particle vertex with three distinct topologies as shown in Fig.\ref{fig:2loop}. Based on our one loop results, we can directly set diagrams (a) and (c) to 0. The diagram (b) is not trivially zero and requires some work. When the contractions are performed and we analyze the tensor structure of the loop integrals, we get two types of contributions: $\epsilon_{ij}k^{i}k^j$ and $\epsilon_{ij} \delta^{ij}$. Both of these reduce to zero for physical polarization of the external graviton.  

Together, one can write the full 2PM corrected contribution, 
\begin{align}
    \label{A_2PM_full}
    i \mathcal{A}_{\rm{2PM}}= i \mathcal{A_{\rm{0PM}}}  & \times (-G M | \omega |)^2 \times \left(-\frac{\omega^2+i \epsilon }{\pi \mu^2} e^{\gamma_E}\right)^{(d-4)} \\ \nn & \left[ -\frac{8}{\delta^2_{\rm{IR}}} + \frac{44}{3}\frac{1}{\delta_{\rm{IR}}} - \frac{214}{105}\frac{1}{\delta_{\rm{UV}}} -\frac{7\pi^2}{6}- \left(\frac{11}{6}\right)^2  - \frac{1777\times2}{14700}- 4 \left(\frac{\pi^2}{8} + \frac{203}{72}\right) \right]
\end{align}
where we have used the shorthand $\delta = (d-4)$. 

We see that we have both IR and UV divergences in our expression. While for the UV divergence we will employ the usual renormalization scheme, which we will describe later, we first discuss the IR divergences. We will show later in Section \ref{sec:BHPT}that these IR divergence are not present in the UV complete BHPT calculation. To resolve this tension, we first note that all the IR divergent terms always exponentiate to a phase factor. Hence, when computing any physical observable at the amplitude squared level, these IR divergences always cancel. Let us first demonstrate explicitly this exponentiation for our scattering amplitude.

The full scattering amplitude is given by, 
\begin{align}
    \frac{ \mathcal{A}}{\mathcal{A_{\rm{0PM}}}} = 1 + \frac{\mathcal{A_{\rm{1PM}}}}{\mathcal{A_{\rm{0PM}}}}+ \frac{\mathcal{A_{\rm{2PM}}}}{\mathcal{A_{\rm{0PM}}}} \equiv \mathcal{\alpha}e^{i \theta}  
\end{align}

Our goal is to show that all the IR divergences are contained in the phase factor $\theta$. We perform the logarithmic operation, $\ln$, on both sides, 
\begin{align}
\label{log_amplitude}
    \ln\left(\frac{ \mathcal{A}}{\mathcal{A_{\rm{0PM}}}}\right) = \ln (\mathcal{\alpha}) + i \theta \equiv \ln \left(1 + \frac{\mathcal{A_{\rm{1PM}}}}{\mathcal{A_{\rm{0PM}}}}+ \frac{\mathcal{A_{\rm{2PM}}}}{\mathcal{A_{\rm{0PM}}}}\right) \sim \frac{\mathcal{A_{\rm{1PM}}}}{\mathcal{A_{\rm{0PM}}}}+\frac{\mathcal{A_{\rm{2PM}}}}{\mathcal{A_{\rm{0PM}}}} - \frac{1}{2} \left(\frac{\mathcal{A_{\rm{1PM}}}}{\mathcal{A_{\rm{0PM}}}}\right)^2
\end{align}
where we used $2GM\omega\sim \epsilon \ll 1$ and kept terms up to second order in $\epsilon$. Therefore to show that all the IR divergences exponentiate, it would suffice to show that the real part of the R.H.S of Eq.~\ref{log_amplitude} does not contain any IR divergence. To this end, we first square the 1PM correction \eqref{A_1PM_full} to get 
\begin{align}
    \left(\frac{\mathcal{A_{\rm{1PM}}}}{\mathcal{A_{\rm{0PM}}}}\right)^2 &= 4  (G M |\omega|)^2 \times \Big[  \pi^2 - \frac{4}{\delta_{\rm{IR}}^2} + \left( \frac{22}{3}- 4\Lambda \right)\frac{1}{\delta_{\rm{IR}}} - \left(\frac{11}{6}\right)^2 + \frac{11}{3}\Lambda - \Lambda^2 \nonumber\\
    &+ 2 i \text{sgn}( \omega)\left(\frac{2}{\delta_{\rm{IR}}} - \frac{11}{6} + \Lambda\right)  \Big]
\end{align}
where, $\Lambda = \ln\left(\frac{\omega^2}{\pi\mu^2}\right)+\gamma_E$
One can similarly decompose the 2PM correction \eqref{A_2PM_full} into real and imaginary parts and obtain, 
\begin{align}
    \frac{\mathcal{A_{\rm{2PM}}}}{\mathcal{A_{\rm{0PM}}}} &= (G M |\omega|)^2 \times\Big[ -\frac{8}{\delta_{\rm{IR}}^2}  + \left( \frac{44}{3} - 8\Lambda  \right)\frac{1}{\delta_{\rm{IR}}} - \frac{214}{105}\frac{1}{\delta_{\rm{UV}}} + \Sigma+ \frac{44}{3}\Lambda - 8\Lambda^2+ 8\pi^2 \nonumber\\
    &+ i\left( 8\pi \text{sgn}(\omega)\frac{1}{\delta_{\rm{IR}}} + 16\pi \text{sgn}(\omega)\right)    \Big]
\end{align}
where, $\Sigma = -\frac{7\pi^2}{16} - \frac{1777\times2}{14700} - \left(\frac{11}{6}\right)^2 - 4 \left(\frac{\pi^2}{8} + \frac{203}{72}\right)$


Summing all the terms in Eq.~\ref{log_amplitude}, we can clearly see that the IR divergence cancels out in the real part of the scattering amplitude and therefore exponentiate. This shows such divergences are spurious and do not contribute to observables. However, since we are doing a matching at the amplitude level, where we also match the phase factor, we need to know precisely which terms can be ignored. Using the recent amplitude level matching calculations \cite{Ivanov:2026icp}, we see that all the IR poles, i.e. powers of $1/\delta_{\text{IR}}$, may be dropped in the $\bar MS$ scheme.

After taking care of the IR divergences, we move on to renormalizing the UV divergences present in the 2PM result \eqref{A_2PM_full}, which simplifies to, 
\begin{align}
    \label{A_2PM_full-2}
     i \mathcal{A}_{\rm{2PM}} &= i \mathcal{A_{\rm{0PM}}}   \times (-G M | \omega |)^2  \left(-\frac{\omega^2+i \epsilon }{\pi \mu^2} e^{\gamma_E}\right)^{(d-4)} \times \left[  - \frac{214}{105}\frac{1}{\delta} -\frac{5\pi^2}{3}  - \frac{656237}{44100} \right]  \\ \nn
     &=i \mathcal{A_{\rm{0PM}}}   \times (G M \omega )^2 \times \left[   - \frac{214}{105} \left(\frac{1}{\delta} +2\log\left(\frac{\omega}{\mu}\right) - \log \pi+\gamma_E  \right)     -\frac{5\pi^2}{3}  - \frac{656237}{44100}             \right]
\end{align}

Power counting in $\epsilon$ tells us that the UV divergence only enters at NNLO order, which therefore necessitates the introduction of renormalized dTLN $\Lambda_{r,\omega^2}^E (\mu)$ in terms of the renormalization scale $\mu$. This finally gives us the 2PM corrected renormalized response function, 
\begin{equation}
\label{renormalised response fn}
    \mathcal{F}_{2\rm{PM}}^E(\omega) = \mathcal{F}^E_{r}(\omega) \times \left(1 + \textcolor{orange} {2GM\omega \pi sgn(\omega)}  - \textcolor{violet}{\frac{428}{105} (GM\omega)^2 \rm{log}(\frac{\omega}{\mu})}  -  (GM\omega)^2 \left[\frac{5 \pi^2}{3}  + \frac{656237}{44100}\right]\right) + (\rm{magnetic})
\end{equation}
We have denoted the logarithm appearing due to the 2-loop UV divergence with \textcolor{violet}{\textit{violet}} color, and in BHPT we will denote such terms in same color for the purpose of easy identification. Since we require our observed scattering amplitude to be independent of the renormalization scale, this induces a running for the $\Lambda_{\omega^2}$ Wilson co-efficient as 
\begin{equation}
\label{log_div-1}
\frac{\partial\Lambda_{r,\omega^2}^{E/B} (\mu)}{\partial~\rm{log} (\mu)} = - \frac{428}{105}\Lambda_{\omega^0}    
\end{equation}
Finally, there is one additional source of logarithmic divergence that we have not considered yet. The \textcolor{violet}{violet} terms in the EFT arise from the tail effect induced NNLO PM correction. One can easily see that such terms are not universal, as they are proportional to the Wilson coefficients of the tidal operators (such as $\Lambda_{\omega^0}$ in eqn. \eqref{log_div-1}) related to the object. The calculation done so far ignores any corrections from the point particle action, assuming they do not interfere with the tidal corrections and are universal terms that cancel out in the matching in the near-far zone factorization paradigm. However, this assumption is broken due to the inherent non-renormalizability of the action. Certain high order (6 loop) diagrams from the point particle action,  as shown by \cite{Saketh:2023bul} in case of BH,  lead to logarithmic UV divergences which can only be absorbed by the tidal Love numbers. Such terms are universal, as it is property of GR itself, and can also be explicitly seen in BHPT calculation for the NS case. However, full computation of these 6-loop diagrams giving rise to these divergences on EFT side is outside the scope of this paper, but we can infer the coefficient of this universal log term assuming consistency of our EFT from the BHPT matching~\footnote{Existence of the universal log term was predicted in \cite{Saketh:2023bul}, which was absent in the NS dTLN calculation of \cite{Chakrabarti:2013lua}. We show in our paper that such terms indeed exist in the NS case as well, providing strong evidence for the universal nature of these logs alluded in \cite{Saketh:2023bul}.}. This allows us to write down the form of the log running including the universal piece as, 
\begin{equation}
\label{log_div-2}
\frac{\partial}{\partial \log(\mu)}\Lambda_{r,\omega^2}^E (\mu) =  \rm{c}  - \frac{428}{105}\Lambda_{\omega^0} 
\end{equation}

where, $\rm{c}$ is a numerical constant that we will derive from BHPT matching. The universal logarithmic terms will be denoted in \textit{\textcolor{red}{red}} colored terms in the EFT and BHPT results. The numerical coefficient for the universal logarithm can be read off from BHPT, which in later section will be shown to be $c = \frac{32}{45}$.

Pulling all the pieces together, we can finally write down the 2-PM corrected response function
\begin{align}
    \label{F_2pm-final}
    \mathcal{F}_{2\rm{PM}}^E(\omega) = &\Biggr\{\Lambda_{\omega^0}^E + \textcolor{orange}{2 (GM\omega)\pi \Lambda_{\omega^0}^E} + (GM\omega)^2
                        \left[  \Lambda_{r,\omega^2}^E (\mu) - \textcolor{violet}{\frac{428}{105}\Lambda^E_{\omega^0}\rm{log}\left(\frac{\omega}{\mu}\right)}  + \textcolor{red}{c~ \rm{log}\left(\frac{\omega}{\mu} \right)  } \right] \\ \nn 
                         &+ \Lambda^E_{\omega^0} (GM\omega)^2 \left[\frac{4 \pi^2}{3}  + \frac{634913}{44100}\right] + \mathfrak{A}_{FZ}     \Biggl\} 
                        + i \biggl\{ (GM\omega) H_{\omega^1}^E + \textcolor{orange}{2 (GM\omega)^2\pi H_{\omega^1}^E}  \biggl\}
\end{align}
We infer then that the near-far zone factorization in the BHPT does not have a one to one correspondence with the factorization of the corrections from tidal operators from the point particle action. It is broken due to the non-renormalizability of the point particle action. Apart from the corrections in the EFT that lead to the universal logarithmic term in red, it can also contribute non-logarithmic finite pieces which are again universal. We include these pieces through $\mathfrak{A}_{FZ}$, which is a placeholder for all such contributions that we have not calculated here. These cannot be determined through consistency alone and a 6-loop calculation arising from the pure point particle action is needed to fix them, which is beyond the scope of this paper. Thus our final result for the TLN at this order is only correct up to an unknown constant which is universal in the sense that it does not depend on the internal properties of the star. 
The full 2-PM corrected scattering amplitude is given by 
\begin{equation}
    \label{full_S_eft}
    i\mathcal{A}_{\rm{2PM}} = -i \frac{\omega^5}{40 \pi M^2_{\text{pl}}} M (GM)^4\mathcal{F}_{2\rm{PM}}^E(\omega)  + (\rm{magnetic})
\end{equation}

We have used this result to match with the BHPT scattering amplitude. Next we turn to the explicit calculation of this scattering amplitude in the full theory. 

\section{Black hole Perturbation theory}
\label{sec:BHPT}

The problem of incoming wave of helicity h scattering off BH background is well studied (see \cite{futterman1987scattering} for review) and we will only briefly recap the relevant parts here. For our purpose of deriving BHPT scattering amplitude results, we will employ the technique developed by Mano-Suzuki-Takasugi (MST method) \cite{Mano:1996vt,Mano:1996gn, Sasaki:2003xr,Mano:1996mf}. This method matches quite well with the EFT approach, since in this method one can systematically obtain a low frequency expansion of the scattering amplitude, which is amenable for matching to the EFT. This, as we saw in the previous sections, is also organized as a low frequency expansion.  

The MST method can be used for a general spin-s perturbation on a Kerr background, which for a GW (i.e. spin-2) perturbation, results in the Teukolsky equation. However, in this paper we will only work with non-rotating compact objects, and the resulting perturbation equation reduces to Regge-Wheeler(RW) equation \cite{regge1957stability} (using a Chandrasekhar transformation). We then solve this using the MST approach following \cite{Casals:2015nja} to obtain the scattering amplitude. The RW equation is given as 
\begin{equation}
    \left[\frac{d^2}{dr_*^2} + (\omega^2 - V(r))\right] X_{\omega \ell } (r)= 0
    \label{eq:RWe}
\end{equation}
with $V(r)$ being defined as
\begin{equation}
    V(r) = \left(1 - \frac{2M}{r} \right) \left[\frac{\ell (\ell+1)}{r^2}  - \frac{6M}{r^3} \right]    
\end{equation}
Where the explicit definition of (radial) RW variable $X(r)$ is given in Appendix \ref{Appn_A}. 
 Furthermore, as our interest is in COs, we must use the reflective boundary condition at the surface of the object. Unlike a BH, any CO will reflect part of the incident wave back from its surface. This reflection depends on the interior properties of the object, and can either be calculated from solving the interior structure (e.g. for NS) or from first principles (e.g. Quantum corrected BH \cite{Abedi:2020ujo, Chakravarti:2023wlc, Datta:2021row}). The reflection coefficient, or reflectivity, then carries the physical information about the constituents and thus can be used to put constraints on hitherto unknown interior microstructure of these COs. We deal with the solution in the stellar interior in Section \ref{sec:Interior soln}. 
 In this section, we focus on the solution in the exterior region of the CO using the MST technique. We have explicitly shown the calculation in Appendix \ref{Appn_A} and produce the final result for the scattering amplitude here. 
 
In case of BH, i.e with purely ingoing boundary condition at horizon, the scattering amplitude reads \cite{Casals:2015nja},
\begin{align}
    \label{S_full_BH}
    S_{BH} = \omega^{-2s}e ^{-2 i \epsilon ln\epsilon} \frac{A^\nu_{C,out}}{A^\nu_{C,in}} \left[ \frac{1+ \beta_\nu \frac{K_{-\nu-1}}{K_{\nu}}  }   {1+ \alpha_\nu \frac{K_{-\nu-1}}{K_{\nu}} } \right] 
\end{align}
The explicit forms of these functions are given in Appendix \ref{Appn_A}. 

However, for the more general case of a compact object, which has a non-zero surface reflectivity, this amplitude is modified to 
\begin{align}
    \label{S_full_ref}
    S=  \omega^{-2s}e ^{-2 i \epsilon ln\epsilon} \frac{A^\nu_{C,out}}{A^\nu_{C,in}} \left[ \frac{1+ \beta_\nu \frac{K_{-\nu-1}}{K_{\nu}} \left(\frac{ \mathfrak{B}_{-\nu-1}}{\mathfrak{B}_{\nu}}\right) }   {1+ \alpha_\nu \frac{K_{-\nu-1}}{K_{\nu}} \left(\frac{ \mathfrak{B}_{-\nu-1}}{\mathfrak{B}_{\nu}}\right)} \right].
\end{align} 
The ratio $\frac{ \mathfrak{B}_{-\nu-1}}{\mathfrak{B}_{\nu}}$, is a function of the physical reflectivity at the surface of the object and can be defined in terms of interior properties of the NS. We therefore relate it to the solution of the RW equation $X$ in the interior of the star, following the technique employed in \cite{Saketh:2024juq}, by defining the matching parameter 
\begin{equation}
    \label{T_defn}
    T =\left.\frac{r}{X}\frac{dX}{dr_{*}}\right|_{r=R_s}
\end{equation} 
where $R_s$ is the radius of the surface of the star. Here, $r_*$ is the tortoise coordinate given by $r_* = r + 2M \ln\left(\frac{r}{2M} - 1\right)$. We will explicitly compute T by solving the RW equation in the interior of the star in Section \ref{sec:Interior soln}. 
 The solution written in terms of $X^{\nu}(r),X^{-\nu-1}(r)$ in the stellar exterior using \eqref{RW_func_defn} and \eqref{RW_defn} allows us to write 
\begin{equation}
    \label{B_from_T}
    \frac{ \mathfrak{B}_{-\nu-1}}{\mathfrak{B}_{\nu}} = \frac{T X^\nu - (r -2M)\frac{dX^\nu}{dr}}{ (r -2M)\frac{dX^{-\nu-1}}{dr} - T X^{-\nu-1}}
\end{equation}
which we can evaluate at the surface of the star $r= R_s$. We will use this to compute eqn.~\eqref{B_from_T} and hence the BHPT scattering amplitude. This may then be expanded order by order in $\omega$ to match to the EFT result.

Before we expand the scattering amplitude order by order in $\epsilon$, another simplification follows from the "near-far factorization" of the scattering amplitude \cite{Ivanov:2022hlo,Ivanov:2022qqt,Saketh:2023bul}. The reasoning of such factorization stems from the physical picture that any wave incident from infinity (in practice, far away from the body) suffers scattering from two sources- the background potential, and the CO surface. Intuitively, one can see it is only the second scattering that contains the information about the tidal response. Further, one can also show that the near and far zone scale as non-integer and integer powers of renormalized angular momentum $\nu$ respectively, barring a tail effect induced log term in far zone \cite{Saketh:2023bul}. Therefore in our work, following the past investigations of TLNs, we will only consider the near zone portion of the scattering amplitude given by 
\begin{align}
    \label{S_NZ_ref}
    S_{NZ}=    \frac{1+ \beta_\nu \frac{K_{-\nu-1}}{K_{\nu}} \left(\frac{ \mathfrak{B}_{-\nu-1}}{\mathfrak{B}_{\nu}}\right) }   {1+ \alpha_\nu \frac{K_{-\nu-1}}{K_{\nu}} \left(\frac{ \mathfrak{B}_{-\nu-1}}{\mathfrak{B}_{\nu}}\right)} .
\end{align}

However, as demonstrated in \cite{Bautista:2023sdf, Saketh:2023bul}, this factorization is not without caveats. The main issue arises in the dTLNs, where the factorization technique along with the matching procedure can correctly recover the logarithmic pieces as shown in the previous section, but the non-log terms may get contributions from the far zone. To correctly assess these terms, one needs to do an expansion of the full scattering amplitude and the corresponding matching, which is outside the scope of this paper.

\subsection{Small frequency expansion}
 We can now do the small-frequency expansion of the near zone scattering amplitude. Here we follow the generic $l$ prescription where we take the physical limit of $l \to n \in \mathbb{N}$ after the expansion at a particular order of $\epsilon$. This gives poles at $l = n$ in the near zone scattering amplitude. In \cite{Bautista:2023sdf}, the authors have shown in the $l = 1, s = 0$ case, such poles also appear in far zone and cancel out in the full scattering amplitude. While an explicit calculation for $s= -2, l = 2$ is not present, we assume the conjecture of \cite{Bautista:2023sdf} that such poles cancel out in full scattering amplitude for all $l$. This is also dictated by the finiteness of the scattering amplitude. We can the write the near zone scattering amplitude as 
 \begin{equation}
     S_{NZ} = 1 - \eta_{20}e^{2i \delta_{20}^{NZ}}
     \label{eq:Snz}
 \end{equation}
The small frequency expansion reads  
\begin{equation}
    \label{BHPT_eta}
    \eta_{20} = 1+ S_1 \epsilon^6 + \textcolor{orange}{\pi S_1} \epsilon^7.
\end{equation}

and

\begin{equation}
    \label{BHPT_delta}
     2\delta^{NZ}_{20} = S_0 \epsilon^5 + \textcolor{orange}{\pi S_0} \epsilon^6 + \left[S_2 - \left( \textcolor{violet}{\frac{107}{105}S_0 \rm{log}(\epsilon) } + \textcolor{red}{\frac{2}{225} \rm{log}(\epsilon)} \right) \right] \epsilon^7
\end{equation}
where
\begin{align}
\begin{split}
    &S_0 = -\frac{2 (6 \mathcal{C}+T_{0}-3)}{75 \left(6 (\mathcal{C}-1) \mathcal{C} \left(4 \mathcal{C}^3+2 \mathcal{C}^2-3\right)-2 \mathcal{C} \left(\mathcal{C} \left(6 \mathcal{C}^2+4 \mathcal{C}+3\right)+3\right) T_{0}-3 \log (1-2 \mathcal{C}) (6 \mathcal{C}+T_{0}-3)\right)}\\
    &S_1 = -\frac{32 \mathcal{C}^4 T_{1}}{25 \left(-6 (\mathcal{C}-1) \mathcal{C} \left(4 \mathcal{C}^3+2 \mathcal{C}^2-3\right)+2 \mathcal{C} \left(\mathcal{C} \left(6 \mathcal{C}^2+4 \mathcal{C}+3\right)+3\right) T_{0}+3 \log (1-2 \mathcal{C}) (6 \mathcal{C}+T_{0}-3)\right)^2}\\
    \end{split}
\end{align}
Here we have expanded the matching quantity $T$ as $T =  T_0 - (iR\omega) T_1  - (R\omega)^2 T_2 +\cdots$, and $\mathcal{C} = \frac{M}{R_s}$ denotes the compactness of the star. We have also not presented the full cumbersome expression of $S_2$ in the main text; the exact expression can be found in the accompanying Mathematica notebook.
We note here that the BHPT encodes the higher PM corrections (violet terms), the universal logarithmic pieces (red terms), and the tail effects (orange terms) that we observed in the EFT. We can now match BHPT results with EFT calculation to obtain the tidal Love and dissipation numbers in terms of the CO properties, encoded in $T$ and $C$. This will be done in Section \ref{sec:Matching}.

In the next section we will now use stellar perturbation theory to explicitly calculate these two quantities $T$ and $C$ in terms of NS and NS+DM interior structure. 

 
\section{Solving the stellar interior }
\label{sec:Interior soln}

To specify the surface reflectivity of a compact object in terms of the properties of the star, we need to solve for the interior structure. Here we will consider the general treatment for a non-rotating CO, representing both NS and DM admixed NS systems. We will treat matter within the NS as a viscous Landau fluid \cite{LandauLifshitzFluid}, specified by some equation of state. We will then solve the resulting Einstein's field equation due to the fluid stress energy tensor order by order in  $\omega$. In this paper, we will solve up to second order in $\omega$, allowing us to access the NNLO dTLN. 

It must be noted that at second order in $\omega$, the Landau fluid model breaks down for viscous fluids and no longer guarantees a subluminal speed of sound in the fluid. Therefore, different fluid models such as the Israel-Stewart model must be used if we want to treat viscosity beyond NLO. In this paper, we will only consider the Landau fluid model. Therefore, for the calculation of $\omega^2$ terms, we drop the viscosity from our model. However, we will keep viscosity related terms in the BHPT scattering amplitude calculation up until  $\omega^6$ order(NLO), to match our results with the previous tidal dissipation number calculated in \cite{Saketh:2024juq}. For the case of the DM as the second fluid, initially we keep our calculation general considering the second fluid to be viscous as well (again up to $\omega^6$ order). However, for our numerical results of dTLN, we ultimately set the viscosity to zero, to remain agnostic towards any particular microscopic particle model. 

We start with the metric for a static spherically symmetric background parametrized by two radial functions $v(r)$ and $\lambda(r)$:
\begin{equation}g^0_{\mu\nu} = \left(\begin{array}{cccc}
 -e^{v(r)} & 0 & 0 & 0 \\
 0 & e^{\lambda(r)} & 0 & 0 \\
 0 & 0 & r^2 & 0 \\
 0 & 0 & 0 & r^2 \sin ^2(\theta) \\ 
\end{array}\right)\end{equation}
where we can relate $\lambda(r)$ to the mass function $m(r)$ inside the CO by $\lambda(r) = \frac{1}{1 - \frac{2 m(r)}{r}}$.
Remaining general for now, the stress energy tensor (SET) is given by two non-interacting viscous Landau fluids,
\begin{equation} 
T_{\mu \nu }= \sum_{i = 1}^2 {T_i}_{\mu \nu } 
\end{equation}
where, the individual SETs are given by,
\begin{equation}
    {T_i}_{\mu \nu }=( \rho_i + p_i ) {u_i}_{\mu} {u_i}_{\nu} + p_i g_{\mu\nu} - 2 \eta_i {P_i}^{\alpha}_{\mu} {P_i}^{\beta}_{\nu} {\sigma_i}_{\alpha\beta} - \zeta_i \nabla_{\alpha} u^{\alpha}_{i} {P_i}_{\mu\nu}\end{equation}
Where \(\eta_i\)  and \(\zeta_i\) are  the shear and bulk viscosity of the $i$th fluid. Furthermore, we have defined 
\[ {P_i}_{\mu\nu}=g_{\mu\nu} + {u_i}_{\mu} {u_i}_{\nu}\]
\[{\sigma_i}_{\mu\nu}=\frac{1}{2} \left( \nabla_\mu {u_i}_{\nu} + \nabla_\nu {u_i}_{\mu} - \frac{2}{3} g_{\mu\nu} \nabla_\alpha u^\alpha \right)\]
For later convenience, we also define $p_{tot}= p_1+p_2$ and $\rho_{tot}=\rho_1+\rho_2$.

Now, we turn our attention to the zeroth order Einstein's equation for a static CO in equilibrium. We see the viscosity terms vanish in the static limit, and given equations of state $\rho_i = \rho_i(p_i)$, we can use the Tolman-Oppenheimer-Volkoff (TOV) equations \cite{PhysRev.55.374} to solve for the background metric:
\begin{equation}
\frac{dm}{dr} = 4\pi \rho_{tot} r^2    
\end{equation}
\begin{equation}
\frac{dp_i}{dr} = -\frac{\left( \rho_i + p_i \right) \left( m + 4\pi r^3 p_{tot} \right)}{r \left( r - 2m \right)}=-\frac{ (\rho_i + p_i )}{2}\frac{d\nu}{dr},
\label{eq:TOV}
\end{equation}
with $p_i(\epsilon) = {p_i}_c -\frac{2\pi}{3}(3 p_{tot}+\rho_{tot})\epsilon^2,  m(\epsilon)= \frac{4 \pi \epsilon^3 \rho_0}{3} $ as initial conditions. Here $\rho_0=\rho_{tot}(0)$, $p_{i_c}$ is the central pressure, and $\epsilon$ is an arbitrarily small number. We can now solve these equations to  determine the background metric $g^0_{\mu\nu}$. For this we integrate outward until we reach a cutoff pressure representing the surface of the star, roughly on the order of $10^{29} \frac{N}{m^2}$. The background metric components can then be determined using the definition of $m(r)$,
\begin{align}
      &\frac{d\lambda}{dr}=\frac{\kappa r^2 \rho_{tot} e^\lambda - e^\lambda +1}{r} , \ \ \ \frac{d\nu}{dr}=\frac{e^\lambda+8\pi r^2p_{tot}}{r}
      \end{align}
We will subsequently solve this system numerically for various choices of baryonic EoS and DM models.

We now consider a metric perturbation $g_{\mu\nu}=g^0_{\mu\nu}+ h_{\mu\nu}$ arising from the interior fluid's motion due to tidal forces. This perturbation assumes spherical symmetry of the background, and thus we can decompose $h_{\mu\nu}$ in the irreducible representation of the SO(3) group, i.e. scalar, vector and tensor spherical harmonics built from $Y_{lm}(\theta, \phi)$. Furthermore, we can distinguish even and odd parity terms. We are specifically interested in tidal deformations, which are even parity terms that at lowest order in the multipole expansion appear at $l=2$. We can choose $m=0$ due to axial symmetry of the background. Working in the Regge-Wheeler gauge, the perturbation can be parameterized as 
\begin{equation}
h_{\mu\nu} = -\left(\frac{r}{R}\right)^2 Y_2(\theta)e^{-i \omega t}\left(\begin{array}{cccc}
\text{H}_0(r)e^{v(r)} & -i \omega\text{H}_1(r) & 0 & 0 \\
-i\omega \text{H}_1(r) & \text{H}_2(r)e^{\lambda(r)} & 0 & 0 \\
0 & 0 & r^2\text{K}(r) & 0 \\
0 & 0 & 0 & r^2\text{K}(r)\sin^2{\theta}
\end{array}\right)
\label{eq:gpert}
\end{equation}
where $R$ is the radius of the star, and we have only considered the dominant $l=2$ mode. 

We can now substitute the perturbed metric into the Einstein field equations to obtain, at leading order in perturbation, $\delta G_{\mu\nu}=\delta(R_{\mu\nu} -\frac{1}{2}Rg_{\mu\nu})=\kappa \delta T_{\mu\nu}$, where, the variation of the Einstein tensor $\delta G_{\mu\nu}$ is given by,
\begin{equation}
\begin{aligned}
\delta G_{\alpha\beta} = \frac{1}{2} \Big( & g_{\alpha\beta} h^{\gamma\delta} R_{\gamma\delta} - h^{}_{\alpha\beta} R - h^{\gamma}{}_{\gamma}{}_{;\alpha}{}_{;\beta} + h^{}_{\beta}{}^{\gamma}{}_{;\alpha}{}_{;\gamma} + h^{}_{\alpha}{}^{\gamma}{}_{;\beta}{}_{;\gamma} \\
& - h^{}_{\alpha\beta}{}^{;\gamma}{}_{;\gamma} + \frac{1}{2}g_{\alpha\beta} g^{\gamma\delta} h^{\epsilon}{}_{\epsilon}{}_{;\gamma}{}_{;\delta} - g_{\alpha\beta} g^{\gamma\delta} h^{}_{\gamma}{}^{\epsilon}{}_{;\delta}{}_{;\epsilon} + \frac{1}{2} g_{\alpha\beta} g^{\gamma\delta} h^{}_{\gamma\delta}{}^{;\epsilon}{}_{;\epsilon} \Big)
\end{aligned}
\end{equation}
where $R_{\mu\nu}$ and $R$ are Ricci tensor and Ricci scalar respectively. We note that covariant derivative $\nabla_\mu$ is defined with respect to the background metric $g^0_{\mu\nu}$. 

To solve the perturbed equation we must obtain the perturbed stress energy tensor $ \delta T_{\mu\nu}$, which in turn requires the perturbations in  velocity $\delta u^{\mu}$, density $\delta \rho$ and pressure $\delta p$. We start by considering that the unperturbed fluid's 4-velocity has no spatial component and from normalization it simply reads, 
\begin{equation}
u^\mu = \left(e^{-\frac{1}{2} v(r)},0,0,0\right)
\end{equation}

We now consider the perturbation of this 4-velocity. To this end, we define the fluid displacement vector $\xi^\mu$ arising from the external tidal field as \cite{lindblom1983quadrupole,thorne1967non}
\begin{equation}
\label{xi_defn}
    \xi^\mu \equiv (\xi^t, \xi^r, \xi^{\theta}, \xi^{\phi}) = e^{-i \omega t} \left(\frac{r}{R}\right)^2 \left( 0, \frac{e^{-\lambda/2} W}{r }Y_2(\theta), -\frac{V}{r^2} Y'_2(\theta), 0 \right)
\end{equation}
where the functions $W(r)$ and $V(r)$ describe the radial and polar displacements respectively. There is no azimuthal displacement or axisymmetric tides($m=0$) due to the non-rotation of the NS. Additionally, we have decomposed the functions in terms of spherical harmonics. We can now calculate the fluid perturbation using \(\delta u^\mu =u^\nu \ \nabla_\nu \xi^\mu\) to obtain:
\begin{equation}
\label{u_perturb1}
\delta u^\mu = -\left(\frac{r}{R}\right)^2 e^{-i \omega t}e^{-v/2}\left(\frac{1}{2} H_0 Y_2(\theta),\frac{i \omega}{r}  e^{-\lambda(r)/2}W\, Y_2(\theta),\frac{i \omega}{r^2}  V \,Y'_2(\theta), 0\right)
\end{equation}

To obtain the density ($\delta \rho$) and pressure($\delta p$) perturbations in the Eulerian frame, we will first consider the co-moving frame. As the timescale for heating due to viscosity far exceeds the orbital timescale, the density perturbation of the fluid can be written down considering only reversible thermodynamic processes \cite{Ghosh:2023vrx, KochanekCoalesingBNS},
\begin{equation}
    \Delta\rho=-(p+\rho)\frac{\Delta \Omega}{\Omega}, \end{equation} 
where $\Omega$ is the volume of a fluid element and is defined as \cite{Saketh:2024juq, thorne1967non}, 
\begin{align}
    \Delta \Omega/ \Omega = \nabla^s_{\mu}\xi^{\mu}+\frac{\delta(g^s)}{2 \,g^s}=  -Y_{2}(\theta)\frac{e^{-i\omega t}}{2 r^2}\left(H_2 + 2K - \frac{12\,V}{r^2} - \frac{2e^{-\lambda/2}[3\,W + rW']}{r^2}\right)
\end{align}
with $g^s$ is the determinant of the induced metric on the spatial hypersurface, and $\nabla^s$ is the covariant derivative in terms of this induced spatial metric. For our diagonal background metric, to first order in perturbation, the induced metric is simply the $3\times3$ spatial part of $g_{\mu\nu}$. Changing the coordinates to the Eulerian coordinate system, we obtain,
\begin{equation}
\delta \rho = \Delta \rho - \xi^r \frac{d\rho}{dr}
\end{equation}
Next, we find the pressure perturbation. We start by defining the adiabatic parameter $\gamma$ relating the fractional variation of baryon number density to that of pressure as,  
\begin{equation}
    \frac{\Delta p}{p} =\gamma \frac{\Delta n}{n}=-\gamma \frac{\Delta v}{v}.
\end{equation}
With this definition, we can write the the Eulerian variation in pressure as,
\begin{equation}
    \delta p = -\gamma \frac{\Delta v}{v}p  - \xi^r\frac{dp}{dr}
\end{equation}
We will also need the perturbation of the energy-momentum conservation equation to solve for all the equations of motion. Again, to linear order in the perturbation, we have
\begin{equation}
\delta (\nabla_{\mu} T^{\mu}_{\nu}) = -h^{\alpha\mu}T_{\alpha\nu;\mu}+g^{\alpha\mu}(T_{\alpha\nu;\mu}-\delta\Gamma^\beta_{\alpha\mu}T_{\beta\nu}-\delta\Gamma^\beta_{\nu\mu}T_{\alpha\beta})=0
\end{equation}
Combining all the results in the first order Einstein's equation $ \delta G_{\mu\nu}=8\pi \delta T_{\mu\nu}$ allows us to eliminate $H_2$ and $H_1$ defined in eqn.~\eqref{eq:gpert} in terms of other variables.

\begin{equation}
   H_2 = H_0 -32 \pi i \omega  e^{-\frac{\nu}{2}}  \left( V_1 \eta_1 + V_2 \eta_2 \right)\end{equation}
and     
\begin{align}
\label{H_1_fin}
H_1(r) &= \frac{-r}{6} \Biggl( 2 H_0 - 2 r K' + K \left( \kappa r^2 p_{tot} e^{\lambda} + e^{\lambda} - 7 \right) - 2\kappa \sum_{j=1}^{2} \left( e^{\frac{\lambda}{2}} W_j(p_j+\rho_j) + 4i\omega e^{\frac{-\nu}{2}}V_j \eta_j \right) \Biggr)
\end{align}
After simplification, the system of equations reduces to 
\begin{align}
\label{master-1}
H_0' &+ \frac{H_0 \left(8 \pi r^2 p_{tot} e^{\lambda}+e^{\lambda}+1\right)}{r} - K' - \frac{2 K}{r} \\ \nn
&- \frac{2 \kappa i \omega e^{-\frac{\nu}{2}} \sum_{j=1}^{2} \eta_j\left((\kappa r^2 p_{tot} e^{\lambda} +e^{\lambda} +1)V_j+r V_j'-e^{\frac{\lambda }{2}} W_j\right)}{r} + \omega ^2 H_1 e^{-\nu(r)} = 0
\end{align}
and
\begin{align}
\label{master-2}
        & \quad H_0' + \frac{ H_0 e^{\lambda} \left(\kappa r^2 p_{tot} - \kappa r^2 \rho_{tot} + 8\right)}{2r}  + K \left(\frac{\kappa r^2 p_{tot} e^{\lambda} - e^{\lambda} - 3}{r}\right) + \frac{K'}{2} \left( \kappa r^2 p_{tot} e^{\lambda} + e^{\lambda} - 3 \right) \nonumber\\
        & - i \kappa \omega \sum_{j=1}^{2} \left\{ \frac{ \eta_j e^{-\frac{\nu}{2}}}{2 r} \left[ 8\left(4 e^{\lambda}+1\right) V_j + V_j' \left(\kappa r^3 e^{\lambda} \rho_{tot} - r e^{\lambda} + r\right) - 2 r^2 V_j'' + 4 e^{\frac{\lambda}{2}} W_j + r^2 e^{\lambda} (H_0 - 4 K) \right] \right. \nn  \\
        & \qquad \left. + r \eta_j e^{\lambda-\nu/2} \left[ \frac{H_0}{2} + K - \frac{6V_j}{r^2} - \frac{e^{-\lambda/2}}{r^2}(3W_j + rW_j') \right] + e^{-\frac{\nu}{2}} \eta_j' \left( e^{\frac{\lambda}{2}} W_j - r V_j' \right) \right\} \nn \\
        & = - r\omega^2 e^{\lambda-\nu} \left( K - \kappa \sum_{j=1}^{2} V_j (\rho_j+p_j) + \text{visc.} \right)
\end{align}
Eqns. \eqref{master-1} and \eqref{master-2} together form a system of master equations for two independent perturbation quantities $H_0$ and $K$. Although $H_1$ appears explicitly in these equations, one can see it can be eliminated using \eqref{H_1_fin}. Here $visc.$ represents terms that contain shear viscosity, $\eta_j$ and the subscripts $1,2$ label the two fluids.   

Since our aim is to match the full theory result to the EFT order by order in $\omega$ to obtain tidal Love numbers up to NNLO,
 we expand the metric perturbations in powers of $\omega$, keeping terms up to second order in $\omega$.   
\begin{equation}
\label{eq:Hexp}
H_0(r)=H^{(0)}(r)+\omega H^{(1)}(r)+\omega ^2H^{(2)}(r)+O(\omega^3)\end{equation} and 
\begin{equation}
\label{eq:Kexp}
K(r)=K^{(0)}(r)+\omega K^{(1)}(r)+\omega ^2K^{(2)}(r)+O(\omega^3)
\end{equation} 
Substituting back these expansions into the master equation, we obtain the differential equations order by order in $\omega$. 
\subsection{Solution at $O(\omega^0)$}

At LO (i.e. $\mathcal{O}(\omega^0)$), the master equations read:

\begin{align}
\label{zerothOrder1}
H^{(0)'} &+ \frac{H^{(0)} \left(8 \pi r^2 p_{tot} e^{\lambda}+e^{\lambda}+1\right)}{r} - K^{(0)'} - \frac{2 K^{(0)}}{r} = 0
\end{align}
and

\begin{align}
\label{ZerothOrder2}
        H^{(0)'} + \frac{ H^{(0)} e^{\lambda} \left(\kappa r^2 p_{tot} - \kappa r^2 \rho_{tot} + 8\right)}{2r}  + K^{(0)} \left(\frac{\kappa r^2 p_{tot} e^{\lambda} - e^{\lambda} - 3}{r}\right) + \frac{K^{(0)'}}{2} \left( \kappa r^2 p_{tot} e^{\lambda} + e^{\lambda} - 3 \right) = 0
\end{align}
These equations match with  \cite{Hinderer:2007mb} up to some differences in convention. 
To solve the LO master equations, one needs to specify boundary conditions \textit{near} the center $(r \to 0)$ of the star to avoid the pole at the center. To this end, we solve the master equations via power series expansion near the origin. We insert Taylor series expansions for all functions appearing in the master equations and solve for the power series expansions of $H_0$ and $K$ around $r=0$.
\begin{align}
H^{(0)}(r) =\ & 1 - \frac{-4 \pi \left(99 p^2 + 36 p \rho + \rho^2\right) + 9 \rho''}{42 \left(3 p + \rho\right)} r^2 \nn \\
& + \frac{1}{7560 (3p + \rho)^2} \bigg[ 
160 \pi^2 (3p + \rho)^2 (414 p^2 + 57 p \rho + 43 \rho^2)\nn \\
& \hspace{4em}
- 24 \pi (201 p - 28 \rho)(3p + \rho) \rho''
- 54 (\rho'')^2
+ 105 (3p + \rho) \rho^{(4)} 
\bigg] r^4 + \mathcal{O}(r^6)
\end{align}
\begin{align}
K^{(0)}(r) =\ & 1 + \frac{8 \pi \left(-6 p^2 + p \rho + \rho^2\right) + 3 \rho''}{14 \left(3 p + \rho\right)} r^2 + \frac{1}{7560 (3p + \rho)^2} \bigg[
800 \pi^2 (3p + \rho)^2 (18 p^2 - 3 p \rho + 11 \rho^2)
\nn \\
 &+ 120 \pi (3p + \rho)(3p + 20 \rho) \rho''
- 54 (\rho'')^2
+ 105 (3p + \rho) \rho^{(4)}
\bigg] r^4 + \mathcal{O}(r^6)
\end{align}

\begin{figure}
    \centering
    \includegraphics[width=0.5\linewidth]{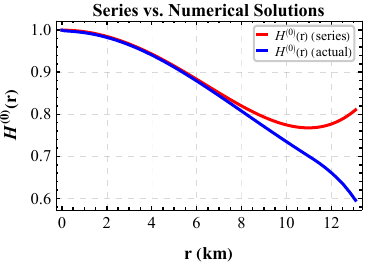}
    \caption{\centering Series expansion (blue) and actual numerical solution (orange)  of $H^{(0)}(r)$ for 2 $M_\odot$ NS .}
    \label{fig:H0}
\end{figure}

Here the pressure, densities, and their derivatives are all evaluated at the origin and we have dropped the $_{tot}$ subscripts for conciseness.
The sixth order terms are too large to be given here and are provided in a supplemental Mathematica notebook. The only constraint on the arbitrary constant term is that $H^{(1)}(0)=K^{(1)}(0)$, thus we choose a value of unity which will not affect our final calculated observables. We show the convergence of the numerical and series solution near the origin for an illustrative case in Fig.~\ref{fig:H0}.

We use these near center solutions for $H_0$ and $K$ as boundary conditions for the numerical integration for the full solution of the master equation, with the other boundary condition at the surface of the star, $r = R_s$. 
The starting point near the center for numerical integration must be chosen to obtain an optimal agreement between the numerical and series solution and their derivatives near the origin. For most NS, starting the numerical integration 10-100 meters from the origin avoids numerical instability and matches smoothly onto the series expansion. In general, with decreasing compactness the matching point moves away from the origin.\footnote{In this case, the pole at the origin could be avoided all together by multiplying by $r$ and making the Cauchy-Euler substitution $r \to e^{s}$. However, we find that the resulting numerical solution does not match the aforementioned series solution near the origin and has a divergent derivative as $r\rightarrow0$. We invoke the regularity of the metric at the origin to disregard this method and instead we use the series solution to begin the numerical integration away from the origin.}

\subsection{Solution at $O(\omega^1)$}
Next, the $O(\omega^1)$ equations are obtained by taking $\omega^2 \to 0$ and substituting the zeroth order solutions in the RHS source terms. We still need to find the static limit of the functions $W$ and $V$ which appear at this order in $\omega$ in our master Eqns.~\eqref{master-1} and~\eqref{master-2}.
Closely following \cite{Saketh:2024juq}, we use the conservation equations $\nabla_{\mu} {T}^{\mu}_{\theta}=\nabla_{\mu} {T}^{\mu}_{r}=0$ for each fluid and take the limit $\omega \to 0$ to obtain alternative expressions for $\delta p$ and $\delta \rho$. Equating these to our previous expressions for these quantities gives us,
\begin{equation}\frac{\Delta v_j}{v_j} = 0,  \ \ \  
    W_j = \frac{r H^{(0)}(p_j+\rho_j)e^{\lambda/2}}{2 p_j'} \text{as}  \ \ \omega \to 0 \end{equation}
This uses the assumption that internal reactions inside the star are very slow compared to $\omega$ even as we approach the static limit. In this case the adiabatic parameter $\gamma$ is given as \cite{Saketh:2024juq}, 
\begin{equation}
\gamma_j = c^2_{ins_j}\frac{p_j+\rho_j}{p_j} \ \ \text{with} \ \ c_{ins_j} = \frac{\partial p_j}{\partial \rho_j} \end{equation}
and the fraction of each particle type in the fluid is held constant in this slow reaction regime.

Note that by substitution of the TOV equation Eq.~\ref{eq:TOV} for each fluid, in static limit we have $W_j =\frac{-r H^{(0)} e^{\lambda/2}}{\nu'}$ indicating the radial fluid perturbations for each fluid are identical. We can then use $\frac{\Delta v_j}{v_j} = 0$ to obtain the static limit of $V_j$ in terms of $W$, $H_0$, and $K$ for both fluids. The expression for the static limit of $V$ can be found in \cite{Saketh:2024juq} and we see that the angular fluid perturbation $V$ is also the same for each fluid.

Now having the source terms for the $\omega^1$ equations, we use the same technique used for the $O(\omega^0)$ result and find a Taylor expansion for $H^{(1)}$ and $K^{(1)}$, defined through Eq.~\ref{eq:Hexp} and Eq.~\ref{eq:Kexp}, around $r=0$ to begin the numerical integration at a point away from the origin. Substituting series expansions for all functions, including for $W$ and $V$ in the static limit, we find that the only constraint for the constant term in the Taylor expansion of $H^{(1)}$ and $K^{(1)}$ is,  
\begin{equation}
K^{(1)}(0)=H^{(1)}(0)+\frac{6e^{-\nu(0)/2}(\eta_1(0)+\eta_2(0))}{3p_{tot}(0)+\rho_{tot}(0)}.
\end{equation}
We can see that, without loss of generality, we can set $H^{(1)}(0)=0$. 

Continuing the expansion to quintic order, we find that only the quadratic and quartic terms in r are nonzero. Here we give the quadratic expansion assuming a non-viscous second fluid. The full general coefficients are given in the supplemental notebook. Again, we drop the $_{tot}$ subscript, and all quantities are evaluated at the origin.
\begin{equation}
\begin{aligned}
H^{(1)}(r)=
&\frac{e^{-\nu/2}}{175\,\kappa (3p+\rho)^4}\Bigg[\, 
\eta_1\Bigl( 50\,\kappa^2(3p+\rho)^4 
+ 15\,\kappa (3p+\rho)(6p+\rho)\,\rho''
- 351\,(\rho'')^2 \\
& + 75\,(3p+\rho)\,\rho^{(4)} \Bigr)  + \frac{75}{2}\,(3p+\rho)\Bigl( \Bigl[ \kappa(3p+\rho)(7p+5\rho) + 6\,\rho'' \Bigr]\eta_1''\\
&- 2(3p+\rho)\,\eta_1^{(4)} \Bigr)
\Bigg]r^2 +{O}(r^4)
\end{aligned}
\end{equation}
\begin{equation}
    \begin{aligned}
      K^{(1)}(r)= & \frac{6e^{-\nu/2}\eta_1}{3p+\rho}+\frac{e^{-\nu/2}}{350\,\kappa (3p+\rho)^4}\Bigg[\,  
\eta_1\Bigl( 50\,\kappa^2 (2p-3\rho)(3p+\rho)^3 \\
&\quad + 9\Bigl( 5\,\kappa (3p+\rho)(17p+5\rho) - 78\,\rho'' \Bigr)\rho'' + 150(3p+\rho)\,\rho^{(4)} \Bigr) \\
& - 150 (3p+\rho) \Bigl( \bigl[ \kappa (3p+\rho)(7p+\rho) - 3\,\rho'' \bigr]\eta_1''  + (3p+\rho)\,\eta_1^{(4)} \Bigr)
\Bigg]r^2+{O}(r^4)
    \end{aligned}
\end{equation}
We can now integrate the $\omega^1$ equations out to arbitrary r using these boundary conditions in the same manner as the LO result. 
\subsection{Solution at $O(\omega^2)$}
The second order equations are only valid for the non-viscous case, $\eta_j \to 0$, as mentioned previously. Therefore, we take all viscous terms (which includes all $\omega^1$ terms) to zero when solving for $H^{(2)}$ and $K^{(2)}$. Similarly to the LO and NLO order calculation, we substitute the static limit perturbations in the $\omega^2$ source terms and solve for the second order metric perturbations, utilizing a Taylor series expansion of the solution near the origin to begin the numerical integration.

At this order, we obtain $H^{(2)}(0)=K^{(2)}(0)$ and again choose unity as the constant term without loss of generality. Similarly, proceeding up to quintic order, only the quadratic and quartic terms in r are nonzero, the latter which are provided in the supplemental notebook.
\begin{equation}
\begin{aligned}
H^{(2)}(r)= 1+\frac{1}{420} \Biggl[\, & -5\,\kappa\,(33p+\rho) \\
& + \frac{54\,e^{-\nu}\,(13p+3\rho)\,\rho''}{\kappa\,(3p+\rho)^3} 
+ \frac{5\,e^{-\nu}\Bigl(-39p+5\rho+18\,e^{\nu}\rho''\Bigr)}{3p+\rho}
\Biggr]r^2+{O}(r^4)
\end{aligned}
\end{equation}
\begin{equation}
\begin{aligned}
K^{(2)}(r)=1+\frac{1}{210}\Biggl[\, & 15\,\kappa\bigl(-2p+\rho\bigr) \\
& + \frac{27\,e^{-\nu}\,(13p+3\rho)\,\rho''}{\kappa\,(3p+\rho)^3}  - \frac{5\,e^{-\nu}\Bigl(9p+\rho-9\,e^{\nu}\rho''\Bigr)}{3p+\rho}
\Biggr]r^2 +{O}(r^4)
\end{aligned}
\end{equation}

To obtain specific numerical solutions, we must supply equations of state for the fluids. We can then integrate the background TOV. equation and solve for the background metric, and the pressure and density profiles for each fluid. Then, with the series expansions for the perturbed metric at the center, we can solve for the interior metric of the entire star up to 2nd order in $\omega$. In the next section, we will describe the observable we compute using the interior solution which will enable us to extract the TLNs. In section \ref{sec:Matching}, we present explicit numerical results.  

\subsection{Surface reflectivity and compactness}

Having obtained the background metric and metric perturbations in the interior, we need to match this solution, or any appropriate function of this solution, to the solution in the exterior. As explained in section~\ref{sec:BHPT}, this is done by computing the coefficients of the small frequency expansion of the matching parameter $T$, along with the star's compactness, which we then use in the BHPT scattering amplitude result eqns.~\eqref{BHPT_eta} and \eqref{BHPT_delta}. 

Here we relate the interior solutions obtained in the previous section, namely the background metric and the metric perturbations to the compactness and the parameter $T$. The compactness ${\cal C} =M/R_s$ can be computed directly from the background metric solution.
To compute $T$, we start by writing the RW variable $X(r)$ introduced in eqn.~\eqref{eq:RWe} 
which can be written inside the star (and at the surface) in terms of the metric perturbation functions $H_0$ and $K$ defined in the previous section \cite{1992PhRvD..46.4289K} ,
\begin{equation} 
\label{X_stellar}
X =\frac{-1}{(n+1)M} \left[ \{-n(n+1)r(r - 3M)\} K + (n+1)r(nr + 3M)e^{-\lambda} H_0 \right]
\end{equation}
and its derivative 
\begin{equation}X' = \frac{1}{-(n+1)M} \left[ n(n+1)\{(n+1)r - 3M\}K + \left\{ -n(n+1)^2 r - 3(n+1)M e^{-\lambda} \right\} H_0 \right]
\end{equation}
where $n=(l-1)(l+2)/2=2$ for $l=2$. 
We now expand  $X$ and its derivative in  $\omega$,  
\[X^{(n)}=X_0^{(n)}+i \omega R\ X_{\omega^1}^{(n)}+(\omega R)^2 X_{\omega^2} ^{(n)}+...\]
where, $n= 0$ denotes $X$ and $n=i$ denotes the $i^{th}$ derivative of RW variable w.r.t. $r$. Substituting the expansion of RW variable in the definition of the matching parameter $T$ given in eqn.~\eqref{T_defn},
\begin{align}
\label{T_interior}
T &= \frac{r}{X} \frac{d  X}{d r_*} \Bigg|_{r=R_S}=\frac{re^{-\lambda}}{X} \frac{d  X}{d r} \Bigg|_{r=R_S} \\ \nn
    &=re^{-\lambda}\left[\frac{X_0'}{X_0}+\frac{-X_1 X_0' +X_0 X_1'}{X_0^2}\omega+(\frac{-X_1 X_1'}{X_0^2}+\frac{X_0'(X_1^2 - X_0 X_2)}{X_0^3}+\frac{X_2'}{X_0})\omega^2 + \cdots \right] \\ \nn
    & \equiv T_0 - (iR\omega) T_1  - (R\omega)^2 T_2 +\cdots
\end{align}

We can then calculate the coefficients in the expansion of $T$ for a specific EoS for NS and NS+DM systems. This will allow us to calculate the `reflectivity' of the stellar surface using eqn.~\eqref{B_from_T}.

\section{Dynamical Tidal response from EFT matching}
\label{sec:Matching}

In the previous sections, we have derived the master equations for stellar perturbations and have seen how they relate to the surface reflectivity. This surface reflectivity, with the BHPT scattering amplitude expression, can then be matched with the EFT result to obtain the static and dynamic TLNs. This is achieved by matching the scattering amplitude obtained from EFT (with 2-PM corrections), eqn.~\eqref{full_S_eft} and the corresponding result from BHPT eqns.~\eqref{eq:Snz}-\eqref {BHPT_delta}, written in terms of the compactness ${\cal C}$ and reflectivity parameter $T$.  we will now focus on numerically solving NS and NS+DM systems with various equations of state for nuclear matter and DM. We then use the results obtained in the previous sections to obtain tidal Love numbers for these choices. 
We will present specific results in the form of rescaled tidal Love numbers, given by 
\begin{align}
        &k^e_2 = \frac{3\mathcal{C}^5}{2}\Lambda_{\omega^0}^E \ \ \ \nu^e_2 = \frac{3\mathcal{C}^6}{2}\Lambda_{\omega^1}^E    \ \ \ \kappa^e_2= \frac{3\mathcal{C}^7}{2}\Lambda_{\omega^2}^E
\end{align}
These rescaled TLNs separate the contributions from compactness and internal structure on the total tidal deformability.

\subsection{Neutron stars admixed with fermionic dark matter}
In the case of a NS, the exact description of interior nuclear matter remains unknown; however, various proposed baryonic EoS have been theorized and have been tabulated in the CompOSE database \cite{stypel_2015_compose}.  To facilitate the numerical solving process, we  fit a piecewise polytropic EoS $(\rho=K p^\Gamma)$ to these EoS. We find that at low pressures, a simple polytrope easily fits the data. At greater pressure and density, it is necessary to fit a sum of polytropes $\rho = K_1\,p^{\Gamma_1}+K_2\,p^{\Gamma_2}$ to obtain an accurate fit of the data. All the equations of state we consider are for low temperature stars, $10^5 - 10^7 \,\text{Kelvin}$. This is appropriate as the vast majority of merging neutron stars are mature enough to have cooled down to this temperature range.

For baryonic matter, the shear viscosity predominantly arises from electron-electron scattering \cite{ShterninShearViscosityNS} and is approximated by the fitting formula \cite{LindblomViscosityNS}

\begin{equation}
    \eta = 270 \frac{\rho^2}{T^2} ~~ (\mathrm{m^{-1}})
\end{equation}
where T is the temperature of the NS. Specifically, we will take $T=10^5 ~\mathrm{K}$ in the following viscosity calculations.

Next we consider the dark matter EoS. We start our discussions with fermionic DM \cite{Arguelles:2023nlh,Profumo:2019ujg} which corresponds to the traditional cold DM (CDM) paradigm. We model the dark matter EoS as a single polytrope, and will only consider non-viscous dark matter for simplicity. We will treat the fermionic DM as a non-relativistic Fermi gas, valid for large particle masses. A simple assumption is to use a DM EoS analogous to that of normal matter in a white dwarf star. This is given by the following parametric equations~\cite{LeungTidalDeform, Oppenheimer:1939ue},

\begin{align}
&\rho = K_{DM} \lbrack \sinh(t)-t \rbrack\\
&p=\frac{K_{DM}}{3} \lbrack \sinh(t)-8\sinh(\frac{t}{2})+3t \rbrack
\end{align}
where $K_{DM}$ is given by:
\[K_{DM} = \left(\frac{M_{\rm{DM}}^4}{32 \pi^2 \hbar^3}\right)\] 
and $M_{\rm{DM}}$ is the mass of the DM particle. For the range of densities and pressures in a NS, one can rewrite these equations as a polytopic EoS in $G=c=1$ units:
\begin{equation}
    \rho=0.00752 \left( \frac{M_{\rm{DM}}}{1\mathrm{GeV}} \right)^\frac{4}{3} p^\frac{2}{3}
\end{equation}

This polytrope is applicable throughout the stars we will consider. However, if the core is very dense and the Fermi energy exceeds $M_{\rm{DM}}$, then in this ultra-relativistic limit we obtain a different polytrope which is not relevant for our purposes. 

\begin{figure}[H]
    \centering
    \includegraphics[width=0.62\linewidth]{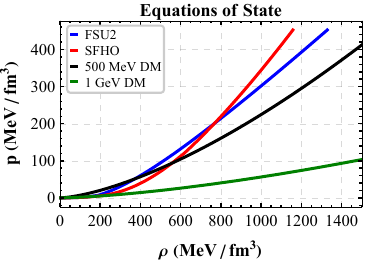}
    \caption{\centering Representative EoS for NS baryonic matter and fermionic DM as a function of energy density. Blue and red curves represent FSU2 and SFHO baryonic EoS while the black and green curves represent EoS for Fermionic DM fluids with particle masses of 500 MeV and 1 GeV respectively.}
    \label{fig:EOSgraphs}
\end{figure}


\begin{figure}[tp]
    \centering
    \begin{subfigure}[t]{0.48\textwidth}
        \centering
            \includegraphics[width=\textwidth]{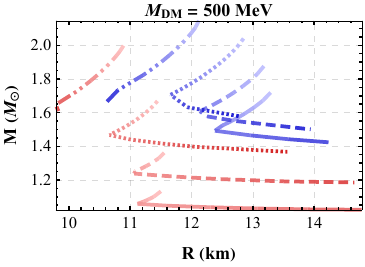}
        \label{MR-500MeV}
    \end{subfigure}
    \hspace{-0.7cm}
    \begin{subfigure}[t]{0.48\textwidth}
        \centering  \includegraphics[ width=\textwidth]{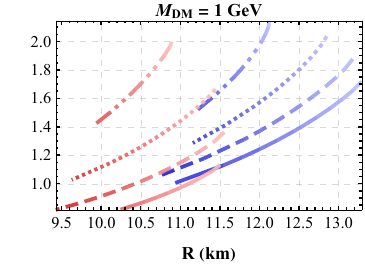}
        \label{MR-1GeV}
    \end{subfigure}
    
    \vspace{-0.4cm}

    \begin{subfigure}[b]           {\textwidth}
        \centering  
         \hspace{1cm}
        \includegraphics[scale=1]{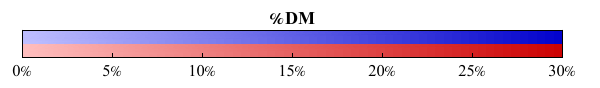}
    \end{subfigure}

    \caption{Mass-radius relationship for fermionic DM admixed neutron star. Two baryonic EoS, FSU2 and SFHO, are denoted with blue and red curves respectively. Each line progresses from light to darker shades, representing lower DM mass fraction ($\%$DM) and higher DM mass fractions respectively.  Solid, dashed, dot-dashed, and dotted lines represent central baryonic densities of 425, 500, 600, and 1000 $\frac{\mathrm{MeV}}{\mathrm{fm^3}}$ respectively.}
    \label{FermionicDM-MR}
\end{figure}
In Fig.~\ref{fig:EOSgraphs} we have plotted some representative EoS of baryonic nuclear matter - FSU2 and SFHO as well as the EoS of fermionic DM for two different masses - 500 MeV and 1 GeV. We focus on fermionic DM particles of mass $\sim O(\mathrm{GeV})$ as this is found to give the richest parameter space consistent with observation. This mass range can be motivated by an attempt to explain matter-antimatter asymmetry by considering a baryon-like dark sector with a corresponding asymmetry \cite{cohen_2010_asymmetric}.

With these EoS, we can solve the background TOV equations to find mass-radius relationships for a family of NS with various cental densities and dark matter fractions. Qualitatively, while the DM gravitationally behaves the same as normal matter, its pressure no longer supports the NS against baryonic matter contraction and vice versa. This effective softening of the star implies that the maximum supported mass for an admixed NS will be lower than that of a pure NS \cite{LeungTidalDeform}. 

This can be seen in Fig.~\ref{FermionicDM-MR}, where we have plotted the mass-radius relation for two different values of admixed fermionic DM mass, $M_{\rm{DM}} = 500 \rm{MeV}$ and $1 \rm{GeV}$ with a varying DM mass fraction ($\%$DM) of the total stellar mass. This is denoted by gradual darkening for each colored curve. For each curve the lightest end denotes the result for pure NS (i.e $0\%$ DM mass). The blue and red curves represent  FSU2 and SFHO EoS respectively, while the each line style denote a particular central baryon density for given EoS.  From the plot, we can see that with increasing DM particle mass($M_{{DM}}$) and DM mass fraction, the allowable total mass of the DM admixed NS drops significantly as previously shown in literature \cite{DM_effect_Ellis}. The range of radii however is less affected, leading to a significant decrease in compactness.

\begin{table}[tp]
    \centering
    \begin{tabular}{lcccccccl}
        \hline
         $\rho_0$&\%DM& M & R (km)&$R_{\rm{DM}}$ (km) & C& $k_{2}^e$&$\nu_{2}^e (\times 10^3)$&  $\kappa^e_2$\\
        \hline
425 & 0 & 1.72 & 13.29 & 0 & 0.19 & 0.087 & 0.032 & 0.16 \\
425 & 1.47 & 1.67 & 13.16 & 8.94 & 0.19 & 0.088 & 0.033 & 0.17 \\
425 & 5.82 & 1.58 & 12.79 & 10.98 & 0.18 & 0.086 & 0.032 & 0.18 \\
425 & 12.11 & 1.49 & 12.3 & 12.38 & 0.18 & 0.082 & 0.03 & 0.2 \\
425 & 21.36 & 1.42 & 11.68 & 14.23 & 0.15 & 0.032 & 0.009 & 0.29 \\
650 & 0 & 2.04 & 12.85 & 0 & 0.24 & 0.052 & 0.00002 & 0.13 \\
650 & 1.61 & 1.99 & 12.74 & 7.95 & 0.23 & 0.053 & 0.021 & 0.14 \\
650 & 5.14 & 1.9 & 12.47 & 9.44 & 0.22 & 0.053 & 0.021 & 0.15 \\
650 & 11.1 & 1.78 & 12.01 & 10.57 & 0.22 & 0.052 & 0.02 & 0.16 \\
650 & 22.27 & 1.63 & 11.19 & 11.93 & 0.2 & 0.037 & 0.013 & 0.19 \\
1000 & 0 & 2.15 & 12.12 & 0 & 0.26 & 0.034 & 0.015 & 0.12 \\
1000 & 1.43 & 2.1 & 12.06 & 6.8 & 0.26 & 0.035 & 0.015 & 0.12 \\
1000 & 4.55 & 2.02 & 11.87 & 8.08 & 0.25 & 0.035 & 0.015 & 0.13 \\
1000 & 9.78 & 1.9 & 11.51 & 9.05 & 0.24 & 0.035 & 0.015 & 0.14 \\
1000 & 19.7 & 1.73 & 10.8 & 10.08 & 0.24 & 0.034 & 0.015 & 0.15 \\
\hline

    \end{tabular}
    \caption{Tabulated numerical values of observables for a 500 MeV fermionic DM mixed with NS with FSU2 equation of state. We consider three distinct values for the central density $\rho_0$ in units of $\frac{\mathrm{MeV}}{\mathrm{fm^3}}$ with increasing mass fraction of DM in each case. }
    \label{tab:FermionDM}
\end{table}
\begin{figure}[tp]
    \centering
 \begin{subfigure}[b]{0.48\textwidth}
        \centering            \includegraphics[trim={0cm 0.5cm 0cm 0cm}, clip,width=\textwidth]{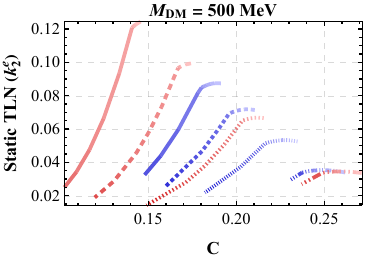}
        \label{500 MeV-ke2}
    \end{subfigure}
        \hspace{-0.7cm}
    \begin{subfigure}[b]{0.48\textwidth}
        \centering
        \includegraphics[trim={0cm 0.5cm 0cm 0cm}, clip,width=\textwidth]{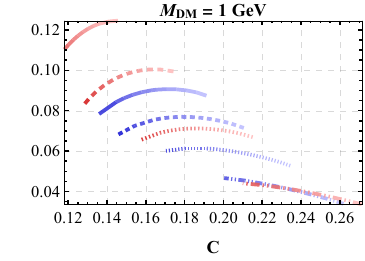}
        \label{1 GeV-ke2}
    \end{subfigure}

\vspace{-0.5cm}
    
    \begin{subfigure}[b]{0.48\textwidth}
        \centering
            \includegraphics[trim={0cm 0.5cm 0cm 0 cm}, clip,width=\textwidth]{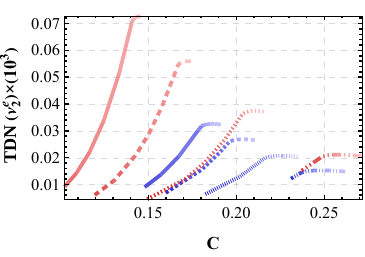}
    \end{subfigure}
        \hspace{-0.7cm}
    \begin{subfigure}[b]{0.48\textwidth}
        \centering
        \includegraphics[trim={0cm 0.5cm 0cm 0cm}, clip,width=\textwidth]{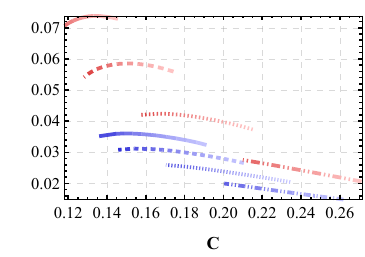}
    \end{subfigure}

\par\smallskip

 \begin{subfigure}[b]{0.48\textwidth}
        \centering
            \includegraphics[width=\textwidth]{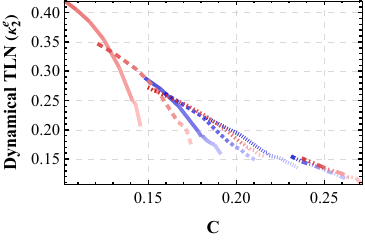}
    \end{subfigure}
        \hspace{-0.7cm}
    \begin{subfigure}[b]{0.48\textwidth}
        \centering
        \includegraphics[width=\textwidth]{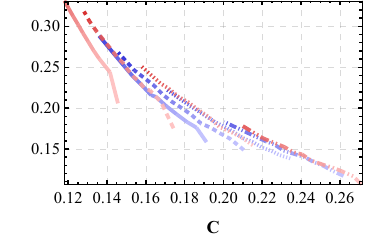}
    \end{subfigure}

    \begin{subfigure}[b]           {\textwidth}
        \centering  
         \hspace{1cm}
         \includegraphics[scale=1]{Plots/colorBar.pdf}
    \end{subfigure}

    \caption{Tidal responses for fermionic DM admixed NS as a function of compactness for two different DM mass, $M_{\rm{DM}} = 500 \rm{MeV}$ and $1 \rm{GeV}$. We have considered FSU and SFHO EoS for baryonic matter, denoted by blue and red curves respectively. Solid, dashed, dot-dashed, and dotted lines represent different central baryonic densities of 425, 500, 600, and 1000 $\frac{\mathrm{MeV}}{\mathrm{fm^3}}$ respectively. For each curve, color shade indicates the DM percentage of total stellar mass, with lighter shades representing lower percentages and darker shades denoting higher percentages.} 
    \label{FermionTLN}
\end{figure}

Next, we determine how the TLNs are affected by the various NS+DM parameters. This requires numerically integrating the perturbative master equations for different DM particle masses and mass fractions to determine how the boundary conditions at the surface of the NS change. These equations mathematically have the same form as in the case of pure normal matter, with the introduction of non-viscous DM simply modifies these equations by changing the total pressure and density, $p_{tot}$ and $\rho_{tot}$. 

In Fig.~\ref{FermionTLN}, we have plotted the static and dynamical TLNs along with TDNs as a function of compactness of the admixed star. As before, we have considered two different EoS for the NS baryonic matter, with blue and red curves representing FSU2 and SFHO EoS respectively, with the various line styles representing different central densities. We have considered two different DM mass of 500MeV and 1 GeV. We have also considered the effect of percentage of total DM mass of total stellar mass, denoted by $\%$DM and indicated by the lighter to darker shading of the curves, with the lightest shade indicating $0\%$ DM contribution, i.e pure NS results. Further, we have tabulated the numerical results in Table~\ref{tab:FermionDM}.

From Fig.~\ref{FermionTLN}, we can see that the static TLN ($k_2^e$) decreases with increasing DM concentration. This can be explained by a dense core forming at the center of the NS as  DM increases, which shifts the mass distribution toward the center of the star. This reduces the amount of mass exposed to the stronger tidal field near the surface, accounting for the lowered tidal deformability. 
The similar decrease in $\nu_2^e$ with increasing percentage of DM is due to the dark matter core increasing the baryon density there and increasing the rigidity of the NS due to gravitational attraction. This decreases the magnitude of fluid displacement within the NS and thus the dissipative energy loss.
Interestingly, the rescaled dynamical TLN instead increases with increasing DM admixture. To interpret this, we recall that $\kappa_2^e$ is correlated with the magnitude of the delayed tidal response solely due to internal structure, after controlling for compactness. As dark matter concentration increases we can understand this increased tidal delay as due to the non-interaction of DM with baryonic matter, removing interactions that would otherwise work to more quickly synchronize a purely baryonic star with the external tidal field.

We also notice that increasing $M_{\rm{DM}}$, which represents a softening of the DM EoS, results in an overall decrease in compactness as qualitatively expected. However, the effect on the static TLN and dissipation number with increasing DM percentage is less pronounced. This is likely due to the compactness reaching such a degree, that adding more DM doesn't significantly overcome the effect of the increased baryonic pressure in order to bring baryonic matter closer to the center. Thus $k^e_2$ and $\nu^e_2$ which are most sensitive to mass distribution within the NS roughly plateau in this regime. 

It is further noted that above a certain critical mass fraction ($\%\rm{DM}_{crit}$), the dense DM core becomes so dominant that the radius of the star abruptly shrinks to $\sim\mathrm{1 km}$ and below. We exclude these unobserved NS from the representative data tables. For fermion masses roughly below $\sim\mathrm{300 MeV}$, the DM instead forms a diffuse halo around the star which drastically reduce the compactness of the NS. GW observations effectively rule out such halos created by even small percentages of fermionic dark matter \cite{barbat_2024_comprehensive} .

\subsection{Neutron Stars Admixed with Bosonic Dark Matter}
\begin{figure}[bp]
    \centering
    \begin{subfigure}[b]{0.48\textwidth}
        \centering
            \includegraphics[width=\textwidth]{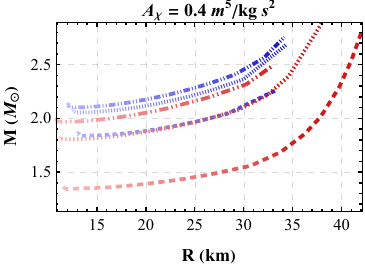}
    \end{subfigure}
        \hspace{-0.7cm}
    \begin{subfigure}[b]{0.48\textwidth}
        \centering
        \includegraphics[width=\textwidth]{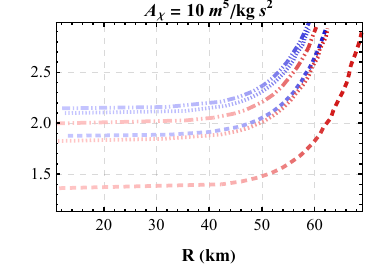}
    \end{subfigure}

        \begin{subfigure}[b] {\textwidth}
        \hspace{1cm}
        \centering  \includegraphics[scale=1]{Plots/colorBar.pdf}
    \end{subfigure}

    \caption{Mass-radius relationship for bosonic DM admixed neutron star. We consider two NS baryonic EoS --- FSU2 and SFHO, given by blue and red respectively. Each curve progresses from lighter to darker shades, indicating a lower or higher DM percentage contribution to the total mass of the system. For each color solid, dashed, and dot-dashed lines represent central baryonic densities of 500, 750, and 1000 $\frac{\mathrm{MeV}}{\mathrm{fm^3}}$ respectively.}
    \label{fig:Bresults1}
\end{figure}
While in the CDM paradigm, DM is modeled as a massive fermionic particle, observational discrepancies have resulted in alternative models like ultralight bosonic DM (see \cite{ferreira2021ultra,Eberhardt:2025caq} for a detailed discussion). In contrast to fermionic dark matter in the GeV range, ultralight bosonic dark matter tends to form diffuse halos around the NS. In our work, we consider the bosonic DM to be given by an ultralight scalar field with quartic self-interaction and given by the equation of state \cite{StubbsBosonicDM,BoehmerHarkoBosonicDM} ,

\begin{equation}
    p=\frac{2\pi l_{DM}}{m_{DM}^3}\rho^2 \equiv A_\chi\rho^2  
\end{equation}
Where $l_{DM}$ is the dark matter particle's scattering length and $m_{DM}$ is its mass.

As before, we treat DM as a non-viscous fluid and numerically integrate the master equations. We will consider the range $0.4< A_\chi<100$ (in $\frac{\mathrm{m}^5}{\mathrm{kg}\,\mathrm{s}^2}$) as this has been shown to produce identifiable effects in admixed NS \cite{StubbsBosonicDM}.
\begin{figure}[t]
    \centering

 \begin{subfigure}[b]{0.48\textwidth}
        \centering
            \includegraphics[trim={0cm 0.5cm 0cm 0cm}, clip,width=\textwidth]{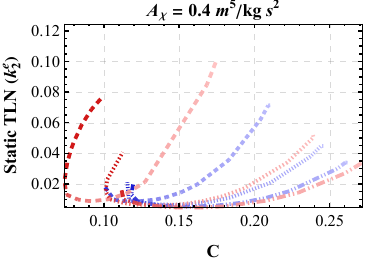}
    \end{subfigure}
        \hspace{-0.7cm}
    \begin{subfigure}[b]{0.48\textwidth}
        \centering
        \includegraphics[trim={0cm 0.5cm 0cm 0cm}, clip,width=\textwidth]{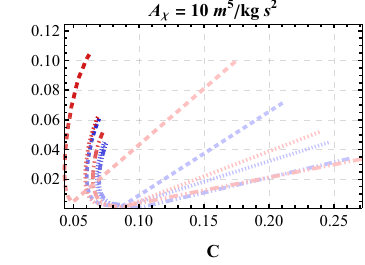}
    \end{subfigure}
    
     \vspace{-0.2cm}
     
    \begin{subfigure}[b]{0.48\textwidth}
        \centering
            \includegraphics[trim={0cm 0.5cm 0cm 0cm}, clip,width=\textwidth]{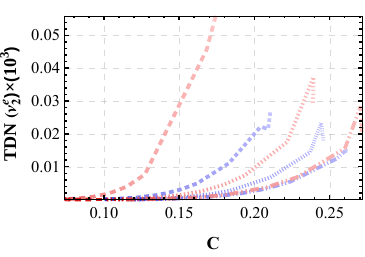}
    \end{subfigure}
        \hspace{-0.7cm}
    \begin{subfigure}[b]{0.48\textwidth}
        \centering
        \includegraphics[trim={0cm 0.5cm 0cm 0cm}, clip,width=\textwidth]{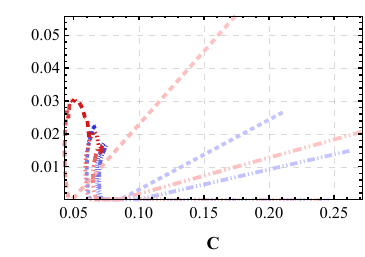}
    \end{subfigure}

\par\smallskip

 \begin{subfigure}[b]{0.48\textwidth}
        \centering        \includegraphics[width=\textwidth]{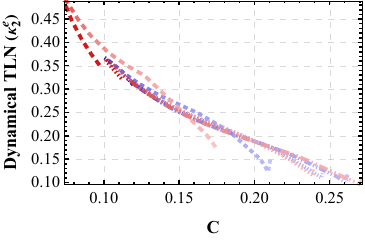}
    \end{subfigure}
        \hspace{-0.7cm}
    \begin{subfigure}[b]{0.48\textwidth}
    \centering  \includegraphics[width=\textwidth]{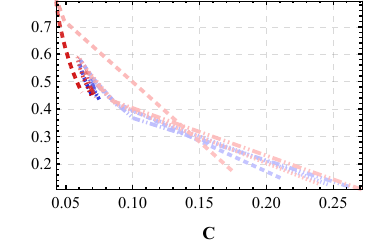}
    \end{subfigure}

    \begin{subfigure}[b]           {\textwidth}
        \centering  
         \hspace{1cm}
        \includegraphics[scale=1]{Plots/colorBar.pdf}
    \end{subfigure}
    
        \caption{Tidal Love numbers for bosonic DM admixed neutron star. FSU and SFHO baryon EoS are blue and red respectively. Each curve progresses from lighter shades (representing lower DM percentage) to darker (higher \%DM) shades. Solid, dashed, and dot-dashed lines represent central baryonic densities of 500, 750, and 1000 $\frac{\mathrm{MeV}}{\mathrm{fm^3}}$ respectively.}
        \label{fig:Bresults2}
\end{figure}
In Fig. \ref{fig:Bresults1} we have plotted the mass-radius relations for the bosonic DM admixed NS for two different values of $A_\chi$ and two different baryonic EoS, FSU2 and SFHO. One can see that owing to its lighter nature, bosonic DM forms a diffuse halo around the NS as expected, compared to fermionic DM admixed NS. It can also be seen explicitly in Table~\ref{tab:Bdm1}, where we have tabulated some representative results.

Next we move on to the calculation of tidal Love and dissipation numbers. We follow the same procedure outlined in the last subsection and produce the tidal responses results in Fig.~\ref{fig:Bresults2}. As before, we consider two different values of $A_\chi$ and two different baryonic EoS, FSU2 and SFHO, given by blue and red curves respectively. Furthermore, in line with the fermionic DM admixed NS, we have chosen three different central baryonic densities given by different line styles. We can see from the plots and the tabulated results that the formation of the larger, diffuse halo reduces its compactness and initially the rescaled static tidal Love number $k_2^e$ decreases as well. However, above a critical amount of DM, $k_{2}^e$ suddenly begins to increase again, signifying an increasing tidal response from the internal structure of the star after controlling for compactness. Note that inverse compactness remains the dominant contribution to tidal deformability, and so the total tidal deformability( $\Lambda_{\omega^0}^E$) still monotonically increases with $\%DM$. Similar behavior is noted for the rescaled dissipation number; however for small enough $A_\chi$ there is no such rebound as the halo grows. This is consistent with the fermion DM case.  In the case of NS with bosonic DM though, the rescaled dynamical TLN instead increases with increasing DM admixture, before reversing only at relatively high dark matter concentrations. The initial increase is consistent with that observed for fermionic DM. The reversal at large DM fraction is likely related  the the extended diffuse halo becoming dominant there. This halo is evidently less responsive to the changing tidal field.

\section{Summary and Outlook}
\label{Conc}

In this paper, we have used a worldline EFT(wEFT) approach to compute dynamical tidal Love numbers for both static neutron stars and neutron stars admixed with dark matter. The TLNs are defined as Wilson coefficients of higher dimensional gauge invariant operators in a point particle wEFT. 
Modeling stars as non-viscous fluids , for the first time we present the tidal Love numbers at Next-to-Next to-Leading order for a variety of dark matter admixed neutron stars in wEFT formalism. We show that loop corrections in the EFT to the TLN operator insertions leads to UV divergences which an be interpreted as the renormalization group evolution for this NNLO tidal Love number. We show that its RG evolution has both a universal and non-universal term. The universal term can be inferred using the consistency of the EFT and we explicitly compute the non-universal term showing that it is proportional to the leading order TLN. 

We note that the tidal Love number computed so far uses the near-far zone factorization framework which is ambiguous up to a universal constant term. To pin down this constant requires a full matching which is beyond the scope of this paper. Few recent papers \cite{Ivanov:2026icp,Bautista:2026qse} have taken the first steps in this direction, performing a full matching to O($\omega^3$). However, for tidal Love numbers we need a matching to at least O($\omega^6$), which will be a non trivial calculation and will be an interesting future work. 
\begin{table}[t]
    \centering
    \begin{tabular}{lcccccccl}
        \hline
         $\rho_0$&\%DM& M & R (km)&$R_{\rm{DM}}$ (km) & C& $k_{2}^e$&$\nu_{2}^e (\times 10^3)$&  $\kappa^e_2$\\
        \hline

500 & 0 & 1.88 & 13.18 & 0 & 0.21 & 0.071 & 0.027 & 0.15 \\
500 & 3.75 & 1.84 & 12.93 & 13.84 & 0.2 & 0.051 & 0.017 & 0.18 \\
500 & 6.5 & 1.85 & 12.83 & 16.92 & 0.16 & 0.02 & 0.0049 & 0.25 \\
500 & 13.04 & 1.94 & 12.7 & 23.38 & 0.12 & 0.0086 & 0.00067 & 0.31 \\
500 & 19.05 & 2.05 & 12.63 & 28.52 & 0.11 & 0.01 & 0.0002 & 0.35 \\
750 & 0 & 2.1 & 12.63 & 0 & 0.25 & 0.045 & 0.000018 & 0.13 \\
750 & 3.7 & 2.05 & 12.39 & 12.82 & 0.24 & 0.038 & 0.014 & 0.13 \\
750 & 10.57 & 2.11 & 12.17 & 19.77 & 0.16 & 0.0069 & 0.00094 & 0.24 \\
750 & 18.32 & 2.26 & 12.06 & 26.54 & 0.13 & 0.0076 & 0.00016 & 0.3 \\
750 & 32.94 & 2.69 & 11.93 & 34.42 & 0.12 & 0.021 & 0.000033 & 0.32 \\
1000 & 0 & 2.15 & 12.12 & 0 & 0.26 & 0.034 & 0.015 & 0.12 \\
1000 & 3.7 & 2.1 & 11.89 & 12.2 & 0.25 & 0.03 & 0.013 & 0.11 \\
1000 & 6.51 & 2.11 & 11.78 & 15.08 & 0.21 & 0.011 & 0.0032 & 0.18 \\
1000 & 11.74 & 2.18 & 11.66 & 20.15 & 0.16 & 0.0052 & 0.00053 & 0.24 \\
1000 & 18.01 & 2.3 & 11.57 & 25.6 & 0.13 & 0.0067 & 0.00012 & 0.28 \\
1000 & 27.67 & 2.56 & 11.48 & 32.02 & 0.12 & 0.013 & 0.000032 & 0.31 \\
        \hline
    \end{tabular}
    \caption{Properties of stars composed of Bosonic DM ($A_\chi=0.4 \frac{\mathrm{m^5}}{\mathrm{kg s^2}}$)  mixed with FSU2 equation of state for non-viscous baryon matter. We consider three distinct central densities $\rho_0$ in units of $\frac{\mathrm{MeV}}{\mathrm{fm^3}}$ with increasing DM fraction. }
    \label{tab:Bdm1}
\end{table}


We have investigated the impact of increasing fraction of Dark Matter on the tidal deformability of the star. 
From the results for the fermionic dark matter admixed neutron stars, we see that with increasing DM particle mass($M_{\rm{DM}}$) and DM mass fraction ($\% \rm{DM}$), the allowable total mass of the NS drops significantly. The range of radii however is less affected, leading to a sizable decrease in compactness. In general, with increasing dark matter admixture, bosonic dark matter forms a larger halo around the neutron star. This reduces its compactness and we discuss in detail the impact on the TLNs in Section \ref{sec:Matching}.

There are various open questions to be explored following our work. Understanding the effect of dynamical TLN on GW waveform is of paramount importance. There are some recent studies on the feasibility of constraining NNLO TLN for BH and neutron stars in next generation GW detectors\cite{Chakraborty:2025wvs, HegadeKR:2024agt, Saes:2025jvr}. In case of purely baryonic NS, the observation prospect of dTLN in next gen detectors has been considered in \cite{Apostolidis:2026qsg}, which shows a significant impact in the GW waveform. With the Wilsonian coefficients for the EFT operators in hand, we can apply this formalism to perform a full analysis for the binary system computing the impact of the TLNs on the Gravitational Wave spectrum. This will tell us which type of binary systems would be the most suitable candidates for phenomenological studies beyond the leading TLNs. This is an important part of our goal of improving the constraints on the microscopic structure of Neutron stars and the accreted Dark Matter halos. Understanding various EoS independent universality relations for dynamical TLN is also another important avenue to investigate in relation to GW astronomy.

Further, in our work we have restricted the calculation to non-rotating stars but most astrophysical Neutron stars have non zero spin. While the BHPT low frequency expansion for Kerr BHs has been recently worked out \cite{Saketh:2023bul}, the case for rotating stars with non-zero reflectivity is still an open problem and is a natural extension of our work. The non-trivial problem however, is solving for the stellar interior with only axial symmetry. 

Finally, in this paper, we have only considered non viscous fluids due to the limitation of the hydrodynamical description. This can be remedied using various prescriptions such as the Israel-Stewart model and would be a natural extension of this work.

\section*{Acknowledgments}
S.M. wants to thank Sumanta Chakraborty, Valerio De Luca, Jordan Wilson-Gerow for insightful discussions on various aspects of Love numbers and tidal dissipation. V.V. is supported by startup funds from the University of South Dakota and by the U.S. Department of Energy, EPSCoR program under contract No. DE-SC0025545. 

\appendix
\section*{Appendix}

\section{MST method for scattering amplitude}
\label{Appn_A}

In this appendix, we elaborate on the MST method used to obtain the scattering amplitude used in the main text. In a general rotating (Kerr) spacetime, the black hole perturbation theory is written in terms of \emph{curvature} perturbation, with the help of Newman-Penrose tetrad formalism. The starting point in this case is the Weyl scalar $\Psi_4 =  -C_{\alpha\beta\gamma\delta} \bar{m}^\alpha l^\beta \bar{m}^\gamma l^\delta$, where, $C_{\alpha\beta\gamma\delta}$ is the Weyl tensor and $l^\alpha, n^\alpha,m^\alpha,\bar{m}^\alpha$ are the null tetrads. The Weyl scalar can then be decomposed into spheroidal harmonics,  $\Psi_4 = e^{-i \omega t} e^{i m \phi} S_\ell^m (\theta) R_{\omega \ell m} (r)$ where the angular part $S_\ell^m (\theta)$ and radial part $R_{\omega \ell m} (r)$ satisfies spheroidal harmonics differential equation and radial Teukolsky equation respectively (see eqn.(9) and (10) of \cite{Sasaki:2003xr}). 
On the other hand, in case of non-rotating BHs, perturbation theory can be written down in terms of \emph{metric} perturbation quantities giving rise to Regge-Wheeler(RW)\cite{regge1957stability} and Zerilli equation \cite{Zerilli:1970se}. Further, for compact objects surface reflectivity is defined through boundary conditions of the RW variable, instead of radial Teukolsky variable which does not have a plane wave representation at the BH horizon. The two formalism however, is connected as it has been shown by Chandrasekhar \cite{chandrasekhar1975equations} that radial Teukolsky equation can be transformed into RW equation in Schwarzschild limit. As in our case, we are only interested in solving non-rotating problem, we will therefore work with RW variable satisfying the equation, 
\begin{equation}
\label{RW_defn}
    \left[\frac{d^2}{dr_*^2} + (\omega^2 - V(r))\right] X_{\omega \ell } (r)= 0
\end{equation}
with, 
\begin{equation}
    V(r) = \left(1 - \frac{2M}{r} \right) \left[\frac{\ell (\ell+1)}{r^2}  - \frac{6M}{r^3} \right]    
\end{equation}
We will now use MST formalism outlined in \cite{Casals:2015nja} to solve the RW equation. In MST formalism, one introduces the renormalized angular momentum $\nu \sim \ell  + \mathcal{O}(\omega^2)$ , and solve the perturbation equation separately in the near-zone ($r_H \leq |r| < \infty$) and far zone ($r_H < |r| \leq \infty$), and then matching in the overlapping intermediate zone $r_H < |r| <\infty$.

In the near-zone, we make a change of variable to $x = \frac{r_H - r}{r_H }$, with the general solution of RW equation in this region can  be written in terms of hypergeometric functions as, 
\begin{align}
    \label{RW_func_defn}
    X = \mathfrak{B}_{\nu} X_0^{\nu}(x) + \mathfrak{B}_{-\nu-1} X_0^{-\nu-1}(x)
\end{align}
where, we have defined, 
\begin{align}
    \label{X_nu}
    X_0^{\nu}(x) = N e^{i \epsilon x} (-x)^{-i \epsilon}  &\left[ \sum^{\infty}_{n = -\infty}  a^\nu_n (1-x)^{n+ \nu+1+i\epsilon} \frac{\Gamma(a)\Gamma(b-a)}{\Gamma(c-a)} \right.  \\ \nn 
    &    \left.  _2F_1(a,c-b,a-b+1; \frac{1}{1-x})   \right]
\end{align}
where,$\epsilon = 2 GM \omega$,  $_2F_1$ is the hypergeometric function and the arguments are given by, $a = -n-\nu +s- i \epsilon, ~~ b= n + \nu +s+1 -i\epsilon,~~ c = 1- 2 i \epsilon$.\footnote{In the expression \eqref{X_nu}, we have corrected a mistype in eqn. 3.23 of \cite{Casals:2015nja}.} Further, $N = \frac{\Gamma(1 - s- 2 i \epsilon)}{\Gamma(-\nu- i \epsilon) \Gamma(1+ \nu- i \epsilon)}$ is a normalization constant. As in this work, we are interested in deriving only up to NNLO TLNs, we have to keep at least up to $\mathcal{O}(\omega^2)$ terms. The renormalized angular momentum $\nu$ and the coefficients $a^\nu_n$ up to this order is given in \cite{Casals:2015nja}, and we have taken the same for our calculation in this paper. Finally the 'reflectivity' terms $\mathfrak{B}_{\nu, -\nu-1}$ needs to be fixed using boundary condition as shown in main text. In case of a BH, $\frac{\mathfrak{B}_{-\nu-1} }{ \mathfrak{B}_{\nu}} = 1$. 

Next we solve eqn. \eqref{RW_defn} at far zone $r_H <  |r| \leq \infty$. In this limit we once again change the variable to $z = \omega r$. In this variable the general solution of RW equation is given in terms of Coulomb functions $X^\nu_C$, and $X^{-\nu-1}_C$, where, 
\begin{align}
    \label{X_C_defn}
    X^{\nu}_{C}(z) = N^{\nu}_{C} e^{-iz} \left(1 - \frac{\bar{\omega}}{z}\right)^{-i\bar{\omega}} (-2iz)^{\nu+1} 
& \sum_{n=-\infty}^{\infty} a^\nu_n \frac{\Gamma(n + \nu + 1 + s - i\bar{\omega})\Gamma(n + \nu + 1 - i\bar{\omega})}{\Gamma(n + \nu + 1 - s + i\bar{\omega})\Gamma(2n + 2\nu + 2)} \\ \nn 
    & \times (2iz)^n M(n + \nu + 1 + i\bar{\omega}, 2n + 2\nu + 2, 2iz)
\end{align}
where, M is confluent hypergeometric function, and the normalization constant $N^\nu_C$ is given by, 
\begin{equation}
    \label{N_nu_C}
    N^{\nu}_{C} = \frac{\Gamma(\nu + 1 - s + i\bar{\omega})\Gamma(\nu + 1 + i\bar{\omega})}{2\Gamma(\nu + 1 + s - i\bar{\omega})\Gamma(\nu + 1 - i\bar{\omega})} e^{i\frac{\pi}{2}(\nu+1)}
\end{equation}
In \cite{Mano:1996gn}, the authors first noted that in the region where both the near zone solution $X_0^\nu$ and the far zone solution $X_C^\nu$ in convergent $(r_H < |r| < \infty)$, satisfy the same analytic properties and therefore can be related to each other proportionally as 
\begin{equation}
    X_0^\nu = K_\nu X^\nu_C
\end{equation}
where

\begin{align} 
\label{Knu_defn}
K_{\nu} =& \frac{e^{i\epsilon}(2\epsilon  )^{s-\nu}2^{-s}
\Gamma(1-s-2i\epsilon)\Gamma(2\nu+2)}{\Gamma(\nu+1-s+i\epsilon) \Gamma(\nu+1+i\epsilon)\Gamma(\nu+1+s+i\epsilon)} \nn \\
& \times \left ( \sum_{n=0}^{\infty} \frac{\Gamma(n+2\nu+1)}{(n)!} \frac{\Gamma(n+\nu+1+s+i\epsilon)}{\Gamma(n+\nu+1-s-i\epsilon)}
\frac{\Gamma(\nu+1+i\epsilon)}{\Gamma(\nu+1-i\epsilon)}\,a^\nu_n\right) \nn \\
&\times \left(\sum_{n=-\infty}^{0} \frac{1}{(-n)! (2\nu+2)_n}\frac{(\nu+1+s-i\epsilon)_n}{(\nu+1-s+i\epsilon)_n} \frac{(\nu+1-i\epsilon)_n}{(\nu+1+i\epsilon)_n}
a^\nu_n\right)^{-1}
\end{align}

Finally, to obtain the scattering amplitude, we would need to connect the far zone solution to the boundary condition at infinity. For this purpose we need to write down $X^{\nu}_{C}(z)$ in terms of incoming and reflected wave at infinity, $\left. X^{\nu}_C\right|_{r=z \to \infty} = A^\nu_{\rm{C,in}} e^{-i \omega r_*}+ A^\nu_{\rm{C,out}} e^{i \omega r_*}$, with the explicit form of them given in eqn. 3.19 and eqn. 3.38 of \cite{Casals:2015nja}. Putting all of these ingredients together, one can obtain the scattering amplitude for the incident perturbation given by RW equation in MST formalism
\begin{align}
    \label{S_full_ref}
    S=  \omega^{-2s}e ^{-2 i \epsilon ln\epsilon} \frac{A^\nu_{C,out}}{A^\nu_{C,in}} \left[ \frac{1+ \beta_\nu \frac{K_{-\nu-1}}{K_{\nu}} \left(\frac{ \mathfrak{B}_{-\nu-1}}{\mathfrak{B}_{\nu}}\right) }   {1+ \alpha_\nu \frac{K_{-\nu-1}}{K_{\nu}} \left(\frac{ \mathfrak{B}_{-\nu-1}}{\mathfrak{B}_{\nu}}\right)} \right].
\end{align}
where,  $\alpha_{\nu} = - i e^{i \pi \nu} \frac{\sin(\nu - s +i \epsilon)}{\sin(\nu+s-i \epsilon)}  $, $\beta_\nu = i e^{i \pi \nu} $.

\newpage
\section{Additional results for TLNs}
Here we present additional results for the fermionic DM admixed neutron stars. 

\begin{figure}[htbp]
    \centering
    \begin{subfigure}[b]{0.45\textwidth}
        \centering
            \includegraphics[width=\textwidth]{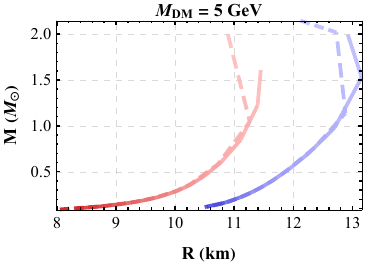}
    \end{subfigure}
    \hfill 
    \begin{subfigure}[b]{0.45\textwidth}
        \centering
        \includegraphics[width=\textwidth]{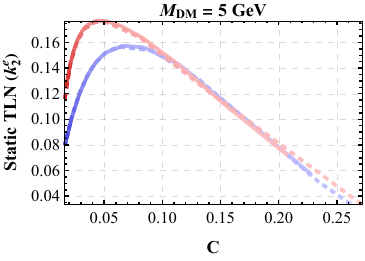}
    \end{subfigure}

\par\smallskip

 \begin{subfigure}[b]{0.45\textwidth}
        \centering
            \includegraphics[width=\textwidth]{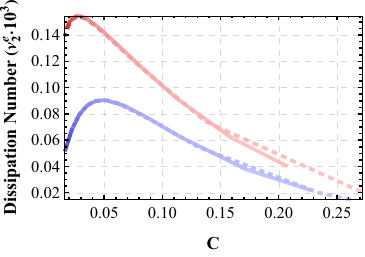}
    \end{subfigure}
    \hfill 
    \begin{subfigure}[b]{0.45\textwidth}
        \centering
        \includegraphics[width=\textwidth]{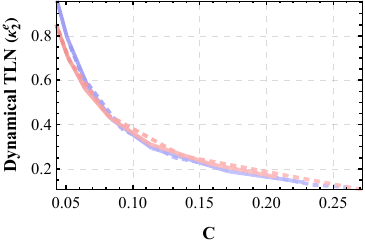}
    \end{subfigure}
    
    \begin{subfigure}[b]           {\textwidth}
        \centering  \includegraphics[scale=1]{Plots/colorBar.pdf}
    \end{subfigure}
    
    \caption{Mass-radius relationship, static and dynamical TLN and tidal dissipation number for fermionic DM admixed neutron star. Two baryonic EoS, FSU2 and SFHO, are denoted with blue and red curves respectively. Each line progresses from light to darker shades, representing lower DM mass fraction ($\%$DM) and higher DM mass fractions respectively.  Solid and dashed lines represent central baryonic densities of 650 and 1000 $\frac{\mathrm{MeV}}{\mathrm{fm^3}}$ respectively.}
\end{figure}

\begin{table}[H]
    \centering
    \begin{tabular}{lcccccccl}
        \hline
         $\rho_0$&\%DM& M & R (km)&$R_{DM}$ (km) & C& $k_{2}^e$&$\nu_{2}^e (\times 10^3)$&  $\kappa^e_2$\\
        \hline
500 & 0.66 & 1.67 & 13.23 & 0.71 & 0.19 & 0.089 & 0.033 & 0.17 \\
500 & 1.08 & 1.48 & 13.18 & 0.61 & 0.17 & 0.11 & 0.04 & 0.19 \\
500 & 1.61 & 1.23 & 13.01 & 0.55 & 0.14 & 0.13 & 0.051 & 0.23 \\
500 & 2.48 & 0.92 & 12.63 & 0.5 & 0.11 & 0.15 & 0.066 & 0.31 \\
800 & 0.67 & 2.01 & 12.82 & 0.64 & 0.23 & 0.054 & 0.021 & 0.14 \\
800 & 1.36 & 1.72 & 13.13 & 0.49 & 0.19 & 0.081 & 0.03 & 0.17 \\
800 & 2.23 & 1.26 & 12.97 & 0.42 & 0.14 & 0.12 & 0.049 & 0.23 \\
800 & 3.13 & 0.95 & 12.6 & 0.4 & 0.11 & 0.14 & 0.064 & 0.31 \\
1500 & 0.79 & 2.12 & 11.9 & 0.57 & 0.26 & 0.031 & 0.014 & 0.12 \\
1500 & 1.58 & 1.89 & 12.92 & 0.39 & 0.22 & 0.063 & 0.024 & 0.15 \\
1500 & 2 & 1.6 & 13.09 & 0.35 & 0.18 & 0.091 & 0.034 & 0.18 \\
1500 & 3.03 & 1.1 & 12.77 & 0.32 & 0.13 & 0.13 & 0.056 & 0.26 \\
    \hline
    \end{tabular}
    \caption{5 GeV fermionic matter with FSU2 equation of state for baryonic matter.}
    \label{tab:placeholder_label}
\end{table}

\pagebreak
\begin{singlespace}
\bibliographystyle{apsrev4-1}
\bibliography{References.bib}

@article{ferreira2021ultra,
  author = "Ferreira, Elisa G. M.",
    title = "{Ultra-light dark matter}",
    eprint = "2005.03254",
    archivePrefix = "arXiv",
    primaryClass = "astro-ph.CO",
    doi = "10.1007/s00159-021-00135-6",
    journal = "Astron. Astrophys. Rev.",
    volume = "29",
    number = "1",
    pages = "7",
    year = "2021"
}

@article{Wang:2026qst,
    author = "Wang, Luohan and Lehner, Luis and Micol, Mait{\'a} and Sturani, Riccardo",
    title = "{Matching Tidal Deformability (Wilson) Coefficients to Black Hole Love Numbers in Higher-Curvature Gravity}",
    eprint = "2604.04259",
    archivePrefix = "arXiv",
    primaryClass = "gr-qc",
    month = "4",
    year = "2026"
}

@article{Oppenheimer:1939ue,
    author = "Oppenheimer, J. R. and Snyder, H.",
    title = "{On Continued gravitational contraction}",
    doi = "10.1103/PhysRev.56.455",
    journal = "Phys. Rev.",
    volume = "56",
    pages = "455--459",
    year = "1939"
}

@article{bildsten1992tidal,
  title={Tidal interactions of inspiraling compact binaries},
  author={Bildsten, Lars and Cutler, Curt},
  journal={Astrophysical Journal, Part 1 (ISSN 0004-637X), vol. 400, no. 1, p. 175-180.},
  volume={400},
  pages={175--180},
  year={1992}
}

@book{poisson2014gravity,
  title={Gravity: Newtonian, post-newtonian, relativistic},
  author={Poisson, Eric and Will, Clifford M},
  year={2014},
  publisher={Cambridge University Press}
}

@article{Hinderer:2007mb,
    author = "Hinderer, Tanja",
    title = "{Tidal Love numbers of neutron stars}",
    eprint = "0711.2420",
    archivePrefix = "arXiv",
    primaryClass = "astro-ph",
    doi = "10.1086/533487",
    journal = "Astrophys. J.",
    volume = "677",
    pages = "1216--1220",
    year = "2008",
    note = "[Erratum: Astrophys.J. 697, 964 (2009)]"
}

@article{Chakraborty:2023zed,
    author = "Chakraborty, Sumanta and Maggio, Elisa and Silvestrini, Michela and Pani, Paolo",
    title = "{Dynamical tidal Love numbers of Kerr-like compact objects}",
    eprint = "2310.06023",
    archivePrefix = "arXiv",
    primaryClass = "gr-qc",
    month = "10",
    year = "2023"
}

@article{Chia:2020yla,
    author = "Chia, Horng Sheng",
    title = "{Tidal deformation and dissipation of rotating black holes}",
    eprint = "2010.07300",
    archivePrefix = "arXiv",
    primaryClass = "gr-qc",
    doi = "10.1103/PhysRevD.104.024013",
    journal = "Phys. Rev. D",
    volume = "104",
    number = "2",
    pages = "024013",
    year = "2021"
}

@article{Porto:2016pyg,
    author = "Porto, Rafael A.",
    title = "{The effective field theorist\textquoteright{}s approach to gravitational dynamics}",
    eprint = "1601.04914",
    archivePrefix = "arXiv",
    primaryClass = "hep-th",
    doi = "10.1016/j.physrep.2016.04.003",
    journal = "Phys. Rept.",
    volume = "633",
    pages = "1--104",
    year = "2016"
}

@article{Mano:1996vt,
    author = "Mano, Shuhei and Suzuki, Hisao and Takasugi, Eiichi",
    title = "{Analytic solutions of the Teukolsky equation and their low frequency expansions}",
    eprint = "gr-qc/9603020",
    archivePrefix = "arXiv",
    reportNumber = "OU-HET-238",
    doi = "10.1143/PTP.95.1079",
    journal = "Prog. Theor. Phys.",
    volume = "95",
    pages = "1079--1096",
    year = "1996"
}

@article{Mano:1996gn,
    author = "Mano, Shuhei and Takasugi, Eiichi",
    title = "{Analytic solutions of the Teukolsky equation and their properties}",
    eprint = "gr-qc/9611014",
    archivePrefix = "arXiv",
    reportNumber = "OU-HET-254",
    doi = "10.1143/PTP.97.213",
    journal = "Prog. Theor. Phys.",
    volume = "97",
    pages = "213--232",
    year = "1997"
}

@article{Sasaki:2003xr,
    author = "Sasaki, Misao and Tagoshi, Hideyuki",
    title = "{Analytic black hole perturbation approach to gravitational radiation}",
    eprint = "gr-qc/0306120",
    archivePrefix = "arXiv",
    doi = "10.12942/lrr-2003-6",
    journal = "Living Rev. Rel.",
    volume = "6",
    pages = "6",
    year = "2003"
}

@article{Cardoso:2017cfl,
    author = "Cardoso, Vitor and Franzin, Edgardo and Maselli, Andrea and Pani, Paolo and Raposo, Guilherme",
    title = "{Testing strong-field gravity with tidal Love numbers}",
    eprint = "1701.01116",
    archivePrefix = "arXiv",
    primaryClass = "gr-qc",
    doi = "10.1103/PhysRevD.95.084014",
    journal = "Phys. Rev. D",
    volume = "95",
    number = "8",
    pages = "084014",
    year = "2017",
    note = "[Addendum: Phys.Rev.D 95, 089901 (2017)]"
}

@article{Brito:2023pyl,
    author = "Brito, Richard and Shah, Shreya",
    title = "{Extreme mass-ratio inspirals into black holes surrounded by scalar clouds}",
    eprint = "2307.16093",
    archivePrefix = "arXiv",
    primaryClass = "gr-qc",
    doi = "10.1103/PhysRevD.108.084019",
    journal = "Phys. Rev. D",
    volume = "108",
    number = "8",
    pages = "084019",
    year = "2023"
}

@article{Maselli:2017cmm,
    author = "Maselli, Andrea and Pani, Paolo and Cardoso, Vitor and Abdelsalhin, Tiziano and Gualtieri, Leonardo and Ferrari, Valeria",
    title = "{Probing Planckian corrections at the horizon scale with LISA binaries}",
    eprint = "1703.10612",
    archivePrefix = "arXiv",
    primaryClass = "gr-qc",
    doi = "10.1103/PhysRevLett.120.081101",
    journal = "Phys. Rev. Lett.",
    volume = "120",
    number = "8",
    pages = "081101",
    year = "2018"
}

@article{Datta:2019epe,
    author = "Datta, Sayak and Brito, Richard and Bose, Sukanta and Pani, Paolo and Hughes, Scott A.",
    title = "{Tidal heating as a discriminator for horizons in extreme mass ratio inspirals}",
    eprint = "1910.07841",
    archivePrefix = "arXiv",
    primaryClass = "gr-qc",
    doi = "10.1103/PhysRevD.101.044004",
    journal = "Phys. Rev. D",
    volume = "101",
    number = "4",
    pages = "044004",
    year = "2020"
}

@article{Nair:2022xfm,
    author = "Nair, Sreejith and Chakraborty, Sumanta and Sarkar, Sudipta",
    title = "{Dynamical Love numbers for area quantized black holes}",
    eprint = "2208.06235",
    archivePrefix = "arXiv",
    primaryClass = "gr-qc",
    doi = "10.1103/PhysRevD.107.124041",
    journal = "Phys. Rev. D",
    volume = "107",
    number = "12",
    pages = "124041",
    year = "2023"
}

@article{Datta:2024vll,
    author = "Datta, Sayak and Brito, Richard and Hughes, Scott A. and Klinger, Talya and Pani, Paolo",
    title = "{Tidal heating as a discriminator for horizons in equatorial eccentric extreme mass ratio inspirals}",
    eprint = "2404.04013",
    archivePrefix = "arXiv",
    primaryClass = "gr-qc",
    month = "4",
    year = "2024"
}

@article{Kain:2021hpk,
    author = "Kain, Ben",
    title = "{Dark matter admixed neutron stars}",
    eprint = "2102.08257",
    archivePrefix = "arXiv",
    primaryClass = "gr-qc",
    doi = "10.1103/PhysRevD.103.043009",
    journal = "Phys. Rev. D",
    volume = "103",
    number = "4",
    pages = "043009",
    year = "2021"
}

@article{Jockel:2023rrm,
    author = "Jockel, C\'edric and Sagunski, Laura",
    title = "{Fermion Proca Stars: Vector Dark Matter Admixed Neutron Stars}",
    eprint = "2310.17291",
    archivePrefix = "arXiv",
    primaryClass = "gr-qc",
    doi = "10.3390/particles7010004",
    journal = "Particles",
    volume = "7",
    number = "1",
    pages = "52--79",
    year = "2024"
}

@article{Cronin:2023xzc,
    author = "Cronin, John and Zhang, Xinyang and Kain, Ben",
    title = "{Rotating dark matter admixed neutron stars}",
    eprint = "2311.07714",
    archivePrefix = "arXiv",
    primaryClass = "gr-qc",
    doi = "10.1103/PhysRevD.108.103016",
    journal = "Phys. Rev. D",
    volume = "108",
    number = "10",
    pages = "103016",
    year = "2023"
}

@article{Diedrichs:2023trk,
    author = {Diedrichs, Robin Fynn and Becker, Niklas and Jockel, C\'edric and Christian, Jan-Erik and Sagunski, Laura and Schaffner-Bielich, J\"urgen},
    title = "{Tidal deformability of fermion-boson stars: Neutron stars admixed with ultralight dark matter}",
    eprint = "2303.04089",
    archivePrefix = "arXiv",
    primaryClass = "gr-qc",
    doi = "10.1103/PhysRevD.108.064009",
    journal = "Phys. Rev. D",
    volume = "108",
    number = "6",
    pages = "064009",
    year = "2023"
}

@article{Saketh:2023bul,
    author = "Saketh, M. V. S. and Zhou, Zihan and Ivanov, Mikhail M.",
    title = "{Dynamical tidal response of Kerr black holes from scattering amplitudes}",
    eprint = "2307.10391",
    archivePrefix = "arXiv",
    primaryClass = "hep-th",
    doi = "10.1103/PhysRevD.109.064058",
    journal = "Phys. Rev. D",
    volume = "109",
    number = "6",
    pages = "064058",
    year = "2024"
}

@article{Ivanov:2022qqt,
    author = "Ivanov, Mikhail M. and Zhou, Zihan",
    title = "{Vanishing of Black Hole Tidal Love Numbers from Scattering Amplitudes}",
    eprint = "2209.14324",
    archivePrefix = "arXiv",
    primaryClass = "hep-th",
    doi = "10.1103/PhysRevLett.130.091403",
    journal = "Phys. Rev. Lett.",
    volume = "130",
    number = "9",
    pages = "091403",
    year = "2023"
}

@article{Ivanov:2022hlo,
    author = "Ivanov, Mikhail M. and Zhou, Zihan",
    title = "{Revisiting the matching of black hole tidal responses: A systematic study of relativistic and logarithmic corrections}",
    eprint = "2208.08459",
    archivePrefix = "arXiv",
    primaryClass = "hep-th",
    doi = "10.1103/PhysRevD.107.084030",
    journal = "Phys. Rev. D",
    volume = "107",
    number = "8",
    pages = "084030",
    year = "2023"
}

@article{Mano:1996mf,
    author = "Mano, Shuhei and Suzuki, Hisao and Takasugi, Eiichi",
    title = "{Analytic solutions of the Regge-Wheeler equation and the postMinkowskian expansion}",
    eprint = "gr-qc/9605057",
    archivePrefix = "arXiv",
    reportNumber = "OU-HET-246",
    doi = "10.1143/PTP.96.549",
    journal = "Prog. Theor. Phys.",
    volume = "96",
    pages = "549--566",
    year = "1996"
}

@article{Saketh:2024juq,
    author = "Saketh, M. V. S. and Zhou, Zihan and Ghosh, Suprovo and Steinhoff, Jan and Chatterjee, Debarati",
    title = "{Investigating tidal heating in neutron stars via gravitational Raman scattering}",
    eprint = "2407.08327",
    archivePrefix = "arXiv",
    primaryClass = "gr-qc",
    doi = "10.1103/PhysRevD.110.103001",
    journal = "Phys. Rev. D",
    volume = "110",
    pages = "103001",
    year = "2024"
}

@article{thorne1967non,
  title={Non-radial pulsation of general-relativistic stellar models. I. Analytic analysis for L>= 2},
  author={Thorne, Kip S and Campolattaro, Alfonso},
  journal={Astrophys. J.},
  volume={149},
  pages={591},
  year={1967}
}

@article{lindblom1983quadrupole,
    author = "Lindblom, L and Detweiler, Steven L.",
    title = "{The quadrupole oscillations of neutron stars}",
    doi = "10.1086/190884",
    journal = "Astrophys. J. Suppl.",
    volume = "53",
    pages = "73--92",
    year = "1983"
}

@article{Casals:2015nja,
    author = "Casals, Marc and Ottewill, Adrian C.",
    title = "{High-order tail in Schwarzschild spacetime}",
    eprint = "1509.04702",
    archivePrefix = "arXiv",
    primaryClass = "gr-qc",
    doi = "10.1103/PhysRevD.92.124055",
    journal = "Phys. Rev. D",
    volume = "92",
    number = "12",
    pages = "124055",
    year = "2015"
}

@article{santos2025observational,
  title={Observational bounds on Dark Matter Admixed Neutron Stars from Gravitational Wave Data},
  author={Santos, Rafael M and Nunes, Rafael C and Coelho, Jaziel G and de Araujo, Jose CN},
  journal={arXiv preprint arXiv:2508.19382},
  year={2025}
}

@article{abbott2018gw170817,
  title={GW170817: Measurements of neutron star radii and equation of state},
  author={Abbott, Benjamin P and Abbott, Richard and Abbott, TD and Acernese, F and Ackley, K and Adams, C and Adams, T and Addesso, P and Adhikari, Rana X and Adya, Vaishali B and others},
  journal={Physical review letters},
  volume={121},
  number={16},
  pages={161101},
  year={2018},
  publisher={APS}
}

@article{LIGOScientific:2014pky,
    author = "Aasi, J. and others",
    collaboration = "LIGO Scientific",
    title = "{Advanced LIGO}",
    eprint = "1411.4547",
    archivePrefix = "arXiv",
    primaryClass = "gr-qc",
    doi = "10.1088/0264-9381/32/7/074001",
    journal = "Class. Quant. Grav.",
    volume = "32",
    pages = "074001",
    year = "2015"
}

@article{VIRGO:2014yos,
    author = "Acernese, F. and others",
    collaboration = "VIRGO",
    title = "{Advanced Virgo: a second-generation interferometric gravitational wave detector}",
    eprint = "1408.3978",
    archivePrefix = "arXiv",
    primaryClass = "gr-qc",
    doi = "10.1088/0264-9381/32/2/024001",
    journal = "Class. Quant. Grav.",
    volume = "32",
    number = "2",
    pages = "024001",
    year = "2015"
}

@article{KAGRA:2020agh,
    author = "Akutsu, T. and others",
    collaboration = "KAGRA",
    title = "{Overview of KAGRA: Calibration, detector characterization, physical environmental monitors, and the geophysics interferometer}",
    eprint = "2009.09305",
    archivePrefix = "arXiv",
    primaryClass = "gr-qc",
    doi = "10.1093/ptep/ptab018",
    journal = "PTEP",
    volume = "2021",
    number = "5",
    pages = "05A102",
    year = "2021"
}

@article{KAGRA:2021vkt,
    author = "Abbott, R. and others",
    collaboration = "KAGRA, VIRGO, LIGO Scientific",
    title = "{GWTC-3: Compact Binary Coalescences Observed by LIGO and Virgo during the Second Part of the Third Observing Run}",
    eprint = "2111.03606",
    archivePrefix = "arXiv",
    primaryClass = "gr-qc",
    reportNumber = "LIGO-P2000318",
    doi = "10.1103/PhysRevX.13.041039",
    journal = "Phys. Rev. X",
    volume = "13",
    number = "4",
    pages = "041039",
    year = "2023"
}

@article{LIGOScientific:2021usb,
    author = "Abbott, R. and others",
    collaboration = "LIGO Scientific, VIRGO",
    title = "{GWTC-2.1: Deep extended catalog of compact binary coalescences observed by LIGO and Virgo during the first half of the third observing run}",
    eprint = "2108.01045",
    archivePrefix = "arXiv",
    primaryClass = "gr-qc",
    reportNumber = "LIGO-P2100063",
    doi = "10.1103/PhysRevD.109.022001",
    journal = "Phys. Rev. D",
    volume = "109",
    number = "2",
    pages = "022001",
    year = "2024"
}

@article{Steinhoff:2016rfi,
    author = "Steinhoff, Jan and Hinderer, Tanja and Buonanno, Alessandra and Taracchini, Andrea",
    title = "{Dynamical Tides in General Relativity: Effective Action and Effective-One-Body Hamiltonian}",
    eprint = "1608.01907",
    archivePrefix = "arXiv",
    primaryClass = "gr-qc",
    doi = "10.1103/PhysRevD.94.104028",
    journal = "Phys. Rev. D",
    volume = "94",
    number = "10",
    pages = "104028",
    year = "2016"
}

@article{Hinderer:2016eia,
    author = "Hinderer, Tanja and others",
    title = "{Effects of neutron-star dynamic tides on gravitational waveforms within the effective-one-body approach}",
    eprint = "1602.00599",
    archivePrefix = "arXiv",
    primaryClass = "gr-qc",
    doi = "10.1103/PhysRevLett.116.181101",
    journal = "Phys. Rev. Lett.",
    volume = "116",
    number = "18",
    pages = "181101",
    year = "2016"
}

@article{Pratten:2021pro,
    author = "Pratten, Geraint and Schmidt, Patricia and Williams, Natalie",
    title = "{Impact of Dynamical Tides on the Reconstruction of the Neutron Star Equation of State}",
    eprint = "2109.07566",
    archivePrefix = "arXiv",
    primaryClass = "astro-ph.HE",
    reportNumber = "LIGO Document P2100307",
    doi = "10.1103/PhysRevLett.129.081102",
    journal = "Phys. Rev. Lett.",
    volume = "129",
    number = "8",
    pages = "081102",
    year = "2022"
}

@article{Ripley:2023lsq,
    author = "Ripley, Justin L. and Hegade K. R., Abhishek and Chandramouli, Rohit S. and Yunes, Nicolas",
    title = "{A constraint on the dissipative tidal deformability of neutron stars}",
    eprint = "2312.11659",
    archivePrefix = "arXiv",
    primaryClass = "gr-qc",
    doi = "10.1038/s41550-024-02323-7",
    journal = "Nature Astron.",
    volume = "8",
    number = "10",
    pages = "1277--1283",
    year = "2024"
}

@article{Ripley:2023qxo,
    author = "Ripley, Justin L. and Hegade K. R., Abhishek and Yunes, Nicolas",
    title = "{Probing internal dissipative processes of neutron stars with gravitational waves during the inspiral of neutron star binaries}",
    eprint = "2306.15633",
    archivePrefix = "arXiv",
    primaryClass = "gr-qc",
    doi = "10.1103/PhysRevD.108.103037",
    journal = "Phys. Rev. D",
    volume = "108",
    number = "10",
    pages = "103037",
    year = "2023"
}

@article{DeLuca:2021ite,
    author = "De Luca, Valerio and Pani, Paolo",
    title = "{Tidal deformability of dressed black holes and tests of ultralight bosons in extended mass ranges}",
    eprint = "2106.14428",
    archivePrefix = "arXiv",
    primaryClass = "gr-qc",
    doi = "10.1088/1475-7516/2021/08/032",
    journal = "JCAP",
    volume = "08",
    pages = "032",
    year = "2021"
}

@article{DeLuca:2022xlz,
    author = "De Luca, Valerio and Maselli, Andrea and Pani, Paolo",
    title = "{Modeling frequency-dependent tidal deformability for environmental black hole mergers}",
    eprint = "2212.03343",
    archivePrefix = "arXiv",
    primaryClass = "gr-qc",
    doi = "10.1103/PhysRevD.107.044058",
    journal = "Phys. Rev. D",
    volume = "107",
    number = "4",
    pages = "044058",
    year = "2023"
}

@article{Nair:2024mya,
    author = "Nair, Sreejith and Chakraborty, Sumanta and Sarkar, Sudipta",
    title = "{Asymptotically de Sitter black holes have nonzero tidal Love numbers}",
    eprint = "2401.06467",
    archivePrefix = "arXiv",
    primaryClass = "gr-qc",
    doi = "10.1103/PhysRevD.109.064025",
    journal = "Phys. Rev. D",
    volume = "109",
    number = "6",
    pages = "064025",
    year = "2024"
}

@article{Barbosa:2025uau,
    author = "Barbosa, Sergio and Brax, Philippe and Fichet, Sylvain and de Souza, Lucas",
    title = "{Running Love numbers and the Effective Field Theory of gravity}",
    eprint = "2501.18684",
    archivePrefix = "arXiv",
    primaryClass = "hep-th",
    doi = "10.1088/1475-7516/2025/07/071",
    journal = "JCAP",
    volume = "07",
    pages = "071",
    year = "2025"
}

@article{Franzin:2024cah,
    author = "Franzin, Edgardo and Frassino, Antonia M. and Rocha, Jorge V.",
    title = "{Tidal Love numbers of static black holes in anti-de Sitter}",
    eprint = "2410.23545",
    archivePrefix = "arXiv",
    primaryClass = "hep-th",
    doi = "10.1007/JHEP12(2024)224",
    journal = "JHEP",
    volume = "12",
    pages = "224",
    year = "2025"
}

@article{Chakravarti:2018vlt,
    author = "Chakravarti, Kabir and Chakraborty, Sumanta and Bose, Sukanta and SenGupta, Soumitra",
    title = "{Tidal Love numbers of black holes and neutron stars in the presence of higher dimensions: Implications of GW170817}",
    eprint = "1811.11364",
    archivePrefix = "arXiv",
    primaryClass = "gr-qc",
    doi = "10.1103/PhysRevD.99.024036",
    journal = "Phys. Rev. D",
    volume = "99",
    number = "2",
    pages = "024036",
    year = "2019"
}

@article{Pereniguez:2021xcj,
    author = "Pere{\~n}iguez, David and Cardoso, Vitor",
    title = "{Love numbers and magnetic susceptibility of charged black holes}",
    eprint = "2112.08400",
    archivePrefix = "arXiv",
    primaryClass = "gr-qc",
    doi = "10.1103/PhysRevD.105.044026",
    journal = "Phys. Rev. D",
    volume = "105",
    number = "4",
    pages = "044026",
    year = "2022"
}

@article{Dey:2020pth,
    author = "Dey, Ramit and Biswas, Shauvik and Chakraborty, Sumanta",
    title = "{Ergoregion instability and echoes for braneworld black holes: Scalar, electromagnetic, and gravitational perturbations}",
    eprint = "2010.07966",
    archivePrefix = "arXiv",
    primaryClass = "gr-qc",
    doi = "10.1103/PhysRevD.103.084019",
    journal = "Phys. Rev. D",
    volume = "103",
    number = "8",
    pages = "084019",
    year = "2021"
}

@article{Chakraborty:2025wvs,
    author = "Chakraborty, Sumanta and De Luca, Valerio and Gualtieri, Leonardo and Pani, Paolo",
    title = "{Dynamical Love numbers of black holes: Theory and gravitational waveforms}",
    eprint = "2507.22994",
    archivePrefix = "arXiv",
    primaryClass = "gr-qc",
    doi = "10.1103/fr3y-s1sz",
    journal = "Phys. Rev. D",
    volume = "112",
    number = "10",
    pages = "104015",
    year = "2025"
}

@article{Kol:2011vg,
    author = "Kol, Barak and Smolkin, Michael",
    title = "{Black hole stereotyping: Induced gravito-static polarization}",
    eprint = "1110.3764",
    archivePrefix = "arXiv",
    primaryClass = "hep-th",
    doi = "10.1007/JHEP02(2012)010",
    journal = "JHEP",
    volume = "02",
    pages = "010",
    year = "2012"
}

@article{LeTiec:2020spy,
    author = "Le Tiec, Alexandre and Casals, Marc",
    title = "{Spinning Black Holes Fall in Love}",
    eprint = "2007.00214",
    archivePrefix = "arXiv",
    primaryClass = "gr-qc",
    doi = "10.1103/PhysRevLett.126.131102",
    journal = "Phys. Rev. Lett.",
    volume = "126",
    number = "13",
    pages = "131102",
    year = "2021"
}

@article{Bhatt:2023zsy,
    author = "Bhatt, Rajendra Prasad and Chakraborty, Sumanta and Bose, Sukanta",
    title = "{Addressing issues in defining the Love numbers for black holes}",
    eprint = "2306.13627",
    archivePrefix = "arXiv",
    primaryClass = "gr-qc",
    reportNumber = "LIGO-P2300180",
    doi = "10.1103/PhysRevD.108.084013",
    journal = "Phys. Rev. D",
    volume = "108",
    number = "8",
    pages = "084013",
    year = "2023"
}

@article{Bhatt:2024yyz,
    author = "Bhatt, Rajendra Prasad and Chakraborty, Sumanta and Bose, Sukanta",
    title = "{Rotating black holes experience dynamical tides}",
    eprint = "2406.09543",
    archivePrefix = "arXiv",
    primaryClass = "gr-qc",
    reportNumber = "LIGO-P2400258",
    doi = "10.1103/PhysRevD.111.L041504",
    journal = "Phys. Rev. D",
    volume = "111",
    number = "4",
    pages = "L041504",
    year = "2025"
}

@article{Katagiri:2024wbg,
    author = "Katagiri, Takuya and Yagi, Kent and Cardoso, Vitor",
    title = "{Relativistic dynamical tides: Subtleties and calibration}",
    eprint = "2409.18034",
    archivePrefix = "arXiv",
    primaryClass = "gr-qc",
    doi = "10.1103/PhysRevD.111.084080",
    journal = "Phys. Rev. D",
    volume = "111",
    number = "8",
    pages = "084080",
    year = "2025"
}

@article{Arguelles:2023nlh,
    author = {Arg{\"u}elles, C. R. and Becerra-Vergara, E. A. and Rueda, J. A. and Ruffini, R.},
    title = "{Fermionic Dark Matter: Physics, Astrophysics, and Cosmology}",
    eprint = "2304.06329",
    archivePrefix = "arXiv",
    primaryClass = "astro-ph.GA",
    doi = "10.3390/universe9040197",
    journal = "Universe",
    volume = "9",
    number = "4",
    pages = "197",
    year = "2023"
}

@article{Profumo:2019ujg,
    author = "Profumo, Stefano and Giani, Leonardo and Piattella, Oliver F.",
    title = "{An Introduction to Particle Dark Matter}",
    eprint = "1910.05610",
    archivePrefix = "arXiv",
    primaryClass = "hep-ph",
    doi = "10.3390/universe5100213",
    journal = "Universe",
    volume = "5",
    number = "10",
    pages = "213",
    year = "2019"
}

@article{Eberhardt:2025caq,
    author = "Eberhardt, Andrew and Ferreira, Elisa G. M.",
    title = "{Ultralight fuzzy dark matter review}",
    eprint = "2507.00705",
    archivePrefix = "arXiv",
    primaryClass = "astro-ph.CO",
    month = "7",
    year = "2025"
}

@article{Mariani:2023wtv,
    author = "Mariani, Mauro and Albertus, Conrado and del Rosario Alessandroni, M. and Orsaria, Milva G. and Perez-Garcia, M. Angeles and Ranea-Sandoval, Ignacio F.",
    title = "{Constraining self-interacting fermionic dark matter in admixed neutron stars using multimessenger astronomy}",
    eprint = "2311.14004",
    archivePrefix = "arXiv",
    primaryClass = "astro-ph.HE",
    doi = "10.1093/mnras/stad3658",
    journal = "Mon. Not. Roy. Astron. Soc.",
    volume = "527",
    number = "3",
    pages = "6795--6806",
    year = "2024"
}

@article{Mukherjee:2025omu,
    author = "Mukherjee, Samanwaya and Aswathi, P. S. and Singha, Chiranjeeb and Ganguly, Apratim",
    title = "{Bose-Einstein Condensate Dark Matter in the Core of Neutron Stars: Implications for Gravitational-wave Observations}",
    eprint = "2506.22353",
    archivePrefix = "arXiv",
    primaryClass = "gr-qc",
    month = "6",
    year = "2025"
}

@article{Saketh:2026trm,
    author = "Saketh, M. V. S. and Ghosh, Suprovo and Andersson, Nils",
    title = "{Dynamical tidal response of neutron stars via scattering amplitudes}",
    eprint = "2606.14405",
    archivePrefix = "arXiv",
    primaryClass = "gr-qc",
    month = "6",
    year = "2026"
}

@article{Apostolidis:2026qsg,
    author = "Apostolidis, Thomas and De Luca, Valerio and Gualtieri, Leonardo and Katagiri, Takuya and Pani, Paolo and Santoni, Luca",
    title = "{Dynamical Tidal Response of Neutron Stars: from Effective Field Theory to Gravitational Waveforms}",
    eprint = "2606.19446",
    archivePrefix = "arXiv",
    primaryClass = "gr-qc",
    month = "6",
    year = "2026"
}

@article{Cipriani:2025tga,
    author = "Cipriani, Lorenzo and Giangrandi, Edoardo and Sagun, Violetta and Doneva, Daniela D. and Yazadjiev, Stoytcho S.",
    title = "{Rapidly spinning dark matter-admixed neutron stars}",
    eprint = "2502.17948",
    archivePrefix = "arXiv",
    primaryClass = "astro-ph.HE",
    doi = "10.1103/qcl7-m5kf",
    journal = "Phys. Rev. D",
    volume = "111",
    number = "12",
    pages = "123005",
    year = "2025"
}

@article{Koehn:2024gal,
    author = "Koehn, Hauke and Giangrandi, Edoardo and Kunert, Nina and Somasundaram, Rahul and Sagun, Violetta and Dietrich, Tim",
    title = "{Impact of dark matter on tidal signatures in neutron star mergers with the Einstein Telescope}",
    eprint = "2408.14711",
    archivePrefix = "arXiv",
    primaryClass = "astro-ph.HE",
    reportNumber = "LA-UR-24-28505",
    doi = "10.1103/PhysRevD.110.103033",
    journal = "Phys. Rev. D",
    volume = "110",
    number = "10",
    pages = "103033",
    year = "2024"
}

@article{Giangrandi:2022wht,
    author = "Giangrandi, Edoardo and Sagun, Violetta and Ivanytskyi, Oleksii and Provid{\^e}ncia, Constan{\c{c}}a and Dietrich, Tim",
    title = "{The Effects of Self-interacting Bosonic Dark Matter on Neutron Star Properties}",
    eprint = "2209.10905",
    archivePrefix = "arXiv",
    primaryClass = "astro-ph.HE",
    doi = "10.3847/1538-4357/ace104",
    journal = "Astrophys. J.",
    volume = "953",
    number = "1",
    pages = "115",
    year = "2023"
}

@article{HegadeKR:2024agt,
    author = "Hegade K. R., Abhishek and Ripley, Justin L. and Yunes, Nicol{\'a}s",
    title = "{Dynamical tidal response of nonrotating relativistic stars}",
    eprint = "2403.03254",
    archivePrefix = "arXiv",
    primaryClass = "gr-qc",
    doi = "10.1103/PhysRevD.109.104064",
    journal = "Phys. Rev. D",
    volume = "109",
    number = "10",
    pages = "104064",
    year = "2024"
}

@article{PhysRev.55.374,
    author = "Oppenheimer, J. R. and Volkoff, G. M.",
    title = "{On massive neutron cores}",
    doi = "10.1103/PhysRev.55.374",
    journal = "Phys. Rev.",
    volume = "55",
    pages = "374--381",
    year = "1939"
}

@article{Goldberger:2004jt,
    author = "Goldberger, Walter D. and Rothstein, Ira Z.",
    title = "{An Effective field theory of gravity for extended objects}",
    eprint = "hep-th/0409156",
    archivePrefix = "arXiv",
    reportNumber = "UCSD-PTH-04-17, CMU-HEP-04-06",
    doi = "10.1103/PhysRevD.73.104029",
    journal = "Phys. Rev. D",
    volume = "73",
    pages = "104029",
    year = "2006"
}

@article{Goldberger:2006bd,
    author = "Goldberger, Walter D. and Rothstein, Ira Z.",
    title = "{Towers of Gravitational Theories}",
    eprint = "hep-th/0605238",
    archivePrefix = "arXiv",
    doi = "10.1142/S0218271806009698",
    journal = "Gen. Rel. Grav.",
    volume = "38",
    pages = "1537--1546",
    year = "2006"
}

@inproceedings{RafieiKarkevandi:2021hcc,
    author = "Rafiei Karkevandi, Davood and Shakeri, Soroush and Sagun, Violetta and Ivanytskyi, Oleksii",
    title = "{Tidal deformability as a probe of dark matter in neutron stars}",
    booktitle = "{16th Marcel Grossmann Meeting on~Recent Developments in Theoretical and Experimental General Relativity, Astrophysics and Relativistic Field Theories}",
    eprint = "2112.14231",
    archivePrefix = "arXiv",
    primaryClass = "astro-ph.HE",
    doi = "10.1142/9789811269776_0307",
    month = "12",
    year = "2021"
}

@article{futterman1987scattering,
  title={Scattering from black holes},
  author={Futterman, J AH and Handler, FA and Matzner, Richard Alfred},
  year={1987},
  publisher={Cambridge University Press, New York, NY}
}

@article{regge1957stability,
  title={Stability of a Schwarzschild singularity},
  author={Regge, Tullio and Wheeler, John A},
  journal={Physical Review},
  volume={108},
  number={4},
  pages={1063},
  year={1957},
  publisher={APS}
}

@article{Bautista:2023sdf,
    author = "Bautista, Yilber Fabian and Bonelli, Giulio and Iossa, Cristoforo and Tanzini, Alessandro and Zhou, Zihan",
    title = "{Black hole perturbation theory meets CFT2: Kerr-Compton amplitudes from Nekrasov-Shatashvili functions}",
    eprint = "2312.05965",
    archivePrefix = "arXiv",
    primaryClass = "hep-th",
    doi = "10.1103/PhysRevD.109.084071",
    journal = "Phys. Rev. D",
    volume = "109",
    number = "8",
    pages = "084071",
    year = "2024"
}

@article{1992PhRvD..46.4289K,
    author = "Kojima, Y.",
    title = "{Equations governing the nonradial oscillations of a slowly rotating relativistic star}",
    doi = "10.1103/PhysRevD.46.4289",
    journal = "Phys. Rev. D",
    volume = "46",
    pages = "4289--4303",
    year = "1992"
}

@article{Ivanov:2026icp,
    author = "Ivanov, Mikhail M. and Li, Yue-Zhou and Parra-Martinez, Julio and Zhou, Zihan",
    title = "{Gravitational Raman Scattering: a Systematic Toolkit for Tidal Effects in General Relativity}",
    eprint = "2602.06951",
    archivePrefix = "arXiv",
    primaryClass = "hep-th",
    reportNumber = "MIT-CTP/6001",
    month = "2",
    year = "2026"
}

@article{chandrasekhar1975equations,
    author = "Chandrasekhar, Subrahmanyan",
    title = "{On the equations governing the perturbations of the Schwarzschild black hole}",
    doi = "10.1098/rspa.1975.0066",
    journal = "Proc. Roy. Soc. Lond. A",
    volume = "343",
    number = "1634",
    pages = "289--298",
    year = "1975"
}

@article{LIGOScientific:2016lio,
    author = "Abbott, B. P. and others",
    collaboration = "LIGO Scientific, Virgo",
    title = "{Tests of general relativity with GW150914}",
    eprint = "1602.03841",
    archivePrefix = "arXiv",
    primaryClass = "gr-qc",
    reportNumber = "LIGO-P1500213",
    doi = "10.1103/PhysRevLett.116.221101",
    journal = "Phys. Rev. Lett.",
    volume = "116",
    number = "22",
    pages = "221101",
    year = "2016",
    note = "[Erratum: Phys.Rev.Lett. 121, 129902 (2018)]"
}

@book{LandauLifshitzFluid,
  title     = {Fluid Mechanics},
  author    = {Landau, L. D. and Lifshitz, E. M.},
  series    = {Course of Theoretical Physics},
  volume    = {6},
  edition   = {2nd},
  year      = {1987},
  publisher = {Butterworth-Heinemann},
  address   = {Oxford},
  isbn      = {978-0750627672}
}

@article{KochanekCoalesingBNS,
    author = "Kochanek, Christopher S.",
    title = "{Coalescing binary neutron stars}",
    reportNumber = "CFA-3361",
    doi = "10.1086/171851",
    journal = "Astrophys. J.",
    volume = "398",
    pages = "234",
    year = "1992"
}

@article{Ghosh:2023vrx,
    author = "Ghosh, Suprovo and Pradhan, Bikram Keshari and Chatterjee, Debarati",
    title = "{Tidal heating as a direct probe of strangeness inside neutron stars}",
    eprint = "2306.14737",
    archivePrefix = "arXiv",
    primaryClass = "gr-qc",
    reportNumber = "LIGO-P2300188",
    doi = "10.1103/PhysRevD.109.103036",
    journal = "Phys. Rev. D",
    volume = "109",
    number = "10",
    pages = "103036",
    year = "2024"
}

@article{Zerilli:1970se,
    author = "Zerilli, Frank J.",
    title = "{Effective potential for even parity Regge-Wheeler gravitational perturbation equations}",
    doi = "10.1103/PhysRevLett.24.737",
    journal = "Phys. Rev. Lett.",
    volume = "24",
    pages = "737--738",
    year = "1970"
}

@article{StubbsBosonicDM,
  title = {Bosonic dark matter dynamics in hybrid neutron stars},
  author = {Buras-Stubbs, Zakary and Lopes, Il\'{\i}dio},
  journal = {Phys. Rev. D},
  volume = {109},
  issue = {4},
  pages = {043043},
  numpages = {13},
  year = {2024},
  month = {Feb},
  publisher = {American Physical Society},
  doi = {10.1103/PhysRevD.109.043043},
  url = {https://link.aps.org/doi/10.1103/PhysRevD.109.043043}
}

@article{BoehmerHarkoBosonicDM,
    author = "Boehmer, C. G. and Harko, T.",
    title = "{Can dark matter be a Bose-Einstein condensate?}",
    eprint = "0705.4158",
    archivePrefix = "arXiv",
    primaryClass = "astro-ph",
    doi = "10.1088/1475-7516/2007/06/025",
    journal = "JCAP",
    volume = "06",
    pages = "025",
    year = "2007"
}

@article{LIGOScientific:2016aoc,
    author = "Abbott, B. P. and others",
    collaboration = "LIGO Scientific, Virgo",
    title = "{Observation of Gravitational Waves from a Binary Black Hole Merger}",
    eprint = "1602.03837",
    archivePrefix = "arXiv",
    primaryClass = "gr-qc",
    reportNumber = "LIGO-P150914",
    doi = "10.1103/PhysRevLett.116.061102",
    journal = "Phys. Rev. Lett.",
    volume = "116",
    number = "6",
    pages = "061102",
    year = "2016"
}

@article{Nissanke:2013fka,
    author = "Nissanke, Samaya and Holz, Daniel E. and Dalal, Neal and Hughes, Scott A. and Sievers, Jonathan L. and Hirata, Christopher M.",
    title = "{Determining the Hubble constant from gravitational wave observations of merging compact binaries}",
    eprint = "1307.2638",
    archivePrefix = "arXiv",
    primaryClass = "astro-ph.CO",
    month = "7",
    year = "2013"
}

@article{LIGOScientific:2017adf,
    author = "Abbott, B. P. and others",
    collaboration = "LIGO Scientific, Virgo, 1M2H, Dark Energy Camera GW-E, DES, DLT40, Las Cumbres Observatory, VINROUGE, MASTER",
    title = "{A gravitational-wave standard siren measurement of the Hubble constant}",
    eprint = "1710.05835",
    archivePrefix = "arXiv",
    primaryClass = "astro-ph.CO",
    reportNumber = "LIGO-P1700296, FERMILAB-PUB-17-472-A-AE",
    doi = "10.1038/nature24471",
    journal = "Nature",
    volume = "551",
    number = "7678",
    pages = "85--88",
    year = "2017"
}

@article{amaro2017laser,
  title={Laser interferometer space antenna},
  author={Amaro-Seoane, Pau and Audley, Heather and Babak, Stanislav and Baker, John and Barausse, Enrico and Bender, Peter and Berti, Emanuele and Binetruy, Pierre and Born, Michael and Bortoluzzi, Daniele and others},
  journal={arXiv preprint arXiv:1702.00786},
  year={2017}
}

@article{LISA:2022kgy,
    author = "Arun, K. G. and others",
    collaboration = "LISA",
    title = "{New horizons for fundamental physics with LISA}",
    eprint = "2205.01597",
    archivePrefix = "arXiv",
    primaryClass = "gr-qc",
    doi = "10.1007/s41114-022-00036-9",
    journal = "Living Rev. Rel.",
    volume = "25",
    number = "1",
    pages = "4",
    year = "2022"
}

@article{ET:2019dnz,
    author = "Maggiore, Michele and others",
    collaboration = "ET",
    title = "{Science Case for the Einstein Telescope}",
    eprint = "1912.02622",
    archivePrefix = "arXiv",
    primaryClass = "astro-ph.CO",
    doi = "10.1088/1475-7516/2020/03/050",
    journal = "JCAP",
    volume = "03",
    pages = "050",
    year = "2020"
}

@article{Goldberger:2009qd,
    author = "Goldberger, Walter D. and Ross, Andreas",
    title = "{Gravitational radiative corrections from effective field theory}",
    eprint = "0912.4254",
    archivePrefix = "arXiv",
    primaryClass = "gr-qc",
    doi = "10.1103/PhysRevD.81.124015",
    journal = "Phys. Rev. D",
    volume = "81",
    pages = "124015",
    year = "2010"
}

@article{Porto:2012as,
    author = "Porto, Rafael A. and Ross, Andreas and Rothstein, Ira Z.",
    title = "{Spin induced multipole moments for the gravitational wave amplitude from binary inspirals to 2.5 Post-Newtonian order}",
    eprint = "1203.2962",
    archivePrefix = "arXiv",
    primaryClass = "gr-qc",
    doi = "10.1088/1475-7516/2012/09/028",
    journal = "JCAP",
    volume = "09",
    pages = "028",
    year = "2012"
}

@article{Chakrabarti:2013lua,
    author = "Chakrabarti, Sayan and Delsate, T{\'e}rence and Steinhoff, Jan",
    title = "{New perspectives on neutron star and black hole spectroscopy and dynamic tides}",
    eprint = "1304.2228",
    archivePrefix = "arXiv",
    primaryClass = "gr-qc",
    month = "4",
    year = "2013"
}

@article{Bautista:2026qse,
    author = "Bautista, Yilber Fabian and Driesse, Mathias and Haddad, Kays and Jakobsen, Gustav Uhre",
    title = "{Gravitational Wave Scattering in Spinless WQFT}",
    eprint = "2602.06125",
    archivePrefix = "arXiv",
    primaryClass = "hep-th",
    reportNumber = "HU-EP-26/05",
    month = "2",
    year = "2026"
}

@article{Saes:2025jvr,
    author = "Saes, Jayana A. and Hegade K. R., Abhishek and Yunes, Nicol{\'a}s",
    title = "{Universal Relations with Dynamical Tides}",
    eprint = "2511.19626",
    archivePrefix = "arXiv",
    primaryClass = "gr-qc",
    month = "11",
    year = "2025"
}

@article{LeungTidalDeform,
    author = "Leung, Kwing-Lam and Chu, Ming-chung and Lin, Lap-Ming",
    title = "{Tidal deformability of dark matter admixed neutron stars}",
    eprint = "2207.02433",
    archivePrefix = "arXiv",
    primaryClass = "astro-ph.HE",
    doi = "10.1103/PhysRevD.105.123010",
    journal = "Phys. Rev. D",
    volume = "105",
    number = "12",
    pages = "123010",
    year = "2022"
}

@article{stypel_2015_compose,
    author = {Typel, S. and Oertel, M. and Kl{\"a}hn, T.},
    title = "{CompOSE CompStar online supernova equations of state harmonising the concert of nuclear physics and astrophysics compose.obspm.fr}",
    eprint = "1307.5715",
    archivePrefix = "arXiv",
    primaryClass = "astro-ph.SR",
    doi = "10.1134/S1063779615040061",
    journal = "Phys. Part. Nucl.",
    volume = "46",
    number = "4",
    pages = "633--664",
    year = "2015"
}

@article{Abedi:2020ujo,
    author = "Abedi, Jahed and Afshordi, Niayesh and Oshita, Naritaka and Wang, Qingwen",
    title = "{Quantum Black Holes in the Sky}",
    eprint = "2001.09553",
    archivePrefix = "arXiv",
    primaryClass = "gr-qc",
    doi = "10.3390/universe6030043",
    journal = "Universe",
    volume = "6",
    number = "3",
    pages = "43",
    year = "2020"
}

@article{Chakravarti:2023wlc,
    author = "Chakravarti, Kabir and Ghosh, Rajes and Sarkar, Sudipta",
    title = "{Formation and stability of area quantized black holes}",
    eprint = "2310.18022",
    archivePrefix = "arXiv",
    primaryClass = "gr-qc",
    doi = "10.1103/PhysRevD.109.046001",
    journal = "Phys. Rev. D",
    volume = "109",
    number = "4",
    pages = "046001",
    year = "2024"
}

@article{Datta:2021row,
    author = "Datta, Sayak and Phukon, Khun Sang",
    title = "{Imprint of black hole area quantization and Hawking radiation on inspiraling binary}",
    eprint = "2105.11140",
    archivePrefix = "arXiv",
    primaryClass = "gr-qc",
    reportNumber = "LIGO-P2100189",
    doi = "10.1103/PhysRevD.104.124062",
    journal = "Phys. Rev. D",
    volume = "104",
    number = "12",
    pages = "124062",
    year = "2021"
}

@article{cohen_2010_asymmetric,
    author = "Cohen, Timothy and Phalen, Daniel J. and Pierce, Aaron and Zurek, Kathryn M.",
    title = "{Asymmetric Dark Matter from a GeV Hidden Sector}",
    eprint = "1005.1655",
    archivePrefix = "arXiv",
    primaryClass = "hep-ph",
    reportNumber = "MCTP-10-18",
    doi = "10.1103/PhysRevD.82.056001",
    journal = "Phys. Rev. D",
    volume = "82",
    pages = "056001",
    year = "2010"
}

@article{DM_effect_Ellis,
    author = {Ellis, John and H{\"u}tsi, Gert and Kannike, Kristjan and Marzola, Luca and Raidal, Martti and Vaskonen, Ville},
    title = "{Dark Matter Effects On Neutron Star Properties}",
    eprint = "1804.01418",
    archivePrefix = "arXiv",
    primaryClass = "astro-ph.CO",
    reportNumber = "CERN-TH-2018-072, KCL-PH-TH/2018-13, KCL-PH-TH-2018-13",
    doi = "10.1103/PhysRevD.97.123007",
    journal = "Phys. Rev. D",
    volume = "97",
    number = "12",
    pages = "123007",
    year = "2018"
}

@article{barbat_2024_comprehensive,
    author = {Barbat, Mikel F. and Schaffner-Bielich, J{\"u}rgen and Tolos, Laura},
    title = "{Comprehensive study of compact stars with dark matter}",
    eprint = "2404.12875",
    archivePrefix = "arXiv",
    primaryClass = "astro-ph.HE",
    doi = "10.1103/PhysRevD.110.023013",
    journal = "Phys. Rev. D",
    volume = "110",
    number = "2",
    pages = "023013",
    year = "2024"
}

@article{LindblomViscosityNS,
    author = "C. Cutler and L. Lindblom",
    title = "{The Effect of Viscosity on Neutron Star Oscillations}",
    doi = "10.1086/165052",
    journal = "Astrophys. J.",
    volume = "314",
    pages = "234",
    year = "1987"
}

@article{ShterninShearViscosityNS,
    author = "Shternin, P. S. and Yakovlev, D. G.",
    title = "{Shear viscosity in neutron star cores}",
    eprint = "0808.2018",
    archivePrefix = "arXiv",
    primaryClass = "astro-ph",
    doi = "10.1103/PhysRevD.78.063006",
    journal = "Phys. Rev. D",
    volume = "78",
    pages = "063006",
    year = "2008"
}

\end{singlespace}

\end{document}